\documentclass[11pt,a4paper]{article}
\usepackage{graphicx}
\usepackage{latexsym}
\usepackage{subfigure}
\usepackage{amsmath}
\usepackage{enumerate}
\usepackage{amssymb}
\usepackage[mathscr]{eucal}
\usepackage{amsthm}
\usepackage{color}
\usepackage[small]{caption}
\newtheorem{thm}{Theorem}[section]

\newtheorem{lem}[thm]{Lemma}

\renewcommand{\d}{\mathrm{d}}
\newcommand{\e}{\mathrm{e}}
\begin{document}

\title{The gap-tooth scheme for homogenization problems}%
\author{Giovanni Samaey and Dirk Roose \\Department of Computer Science, K.U. Leuven \\
Celestijnenlaan 200A, 3001 Leuven, Belgium \and Ioannis G. Kevrekidis \\
Department of Chemical Engineering and PACM\\ Princeton University, Princeton, NJ08544}
\date{\today}
%\address{Celestijnenlaan }
%\email{giovanni.samaey@cs.kuleuven.ac.be}

\maketitle

\begin{abstract}
An important class of problems exhibits smooth
behaviour in space and time on a macroscopic scale,
while only a microscopic evolution law is known.  For such
time-dependent multi-scale problems, an ``equation-free framework''
has been proposed, of which the gap-tooth scheme is an essential component.
The gap-tooth scheme is
designed to approximate a time-stepper for an unavailable macroscopic
equation in a macroscopic domain; it uses
appropriately initialized simulations of the available microscopic
model in a number of small boxes, which cover only a fraction of
the domain.
We analyze the convergence of this
scheme for a parabolic homogenization problem with non-linear
reaction. In this case, the microscopic model is a partial
differential equation with rapidly oscillating coefficients, while
the unknown macroscopic model is
approximated by the homogenized equation. We
show that our method approximates a finite difference scheme of
arbitrary (even) order for
the homogenized equation when we
appropriately constrain the microscopic problem in the boxes. We
illustrate this theoretical result with numerical tests on several
model problems. We also demonstrate that it is possible to obtain
a convergent scheme without constraining the microscopic code, by
introducing buffer regions around the computational boxes.
\end{abstract}

% ----------------------------------------------------------------
\section{Introduction \label{sec:introduction}}

For an important class of multi-scale problems, a separation of
scales exists between the (microscopic, detailed) level of
description of the available model, and the (macroscopic,
continuum) level at which one would like to observe the system.
Consider, for example, a kinetic Monte Carlo model of bacterial
growth \cite{SetGearOthKevr03}. A stochastic model describes the
probability of an individual bacterium to run or ``tumble", based on
the rotation of its flagellae.  Technically, it would be possible
to evolve the detailed model for all space and time, and observe the
macroscopic variables of interest, but this would be prohibitively
expensive.  It is known, however, that, under certain conditions,
one can write a closed deterministic equation for the evolution of
the macroscopic observable (here \emph{bacteria concentration},
the zeroth moment of the evolving distribution) as a
function of macroscopic space and time.

The recently proposed \emph{equation-free framework} \cite{KevrGearHymKevrRunTheo02}
can then be used instead of stochastic time integration in the
entire space-time domain.
This framework is built around the central idea
of a \emph{coarse time-stepper}, which is a time-$\Delta t$ map from
coarse variables to coarse variables.  It consists of the following steps:
(1) {\it lifting}, i.e.~the creation of \emph{appropriate} initial conditions for the microscopic
model; (2) {\it evolution}, using the microscopic model and (possibly) some constraints;
and (3) {\it restriction}, i.e.~the projection of the detailed solution to the macroscopic
``observation" variables.
This coarse time-stepper can subsequently be used as ``input" for a host
time-stepper based algorithms performing macroscopic numerical analysis tasks.
These incude, for example, time-stepper based
bifurcation code to perform bifurcation analysis for the \emph{unavailable} macroscopic
equation \cite{TheoSanSunKevr02,TheoQianKevr00,MakMarKevr02,MakMarPanKevr02}.
A coarse timestepper can also be used in conjunction with
a \emph{projective integration method} to
increase efficiency of time-integration \cite{GearKevr01}.
This approach has
already been used in several applications \cite{SietGrahKevr02,HumKevr02}, and also allows
to do other system level tasks, such as control and optimzization \cite{SietArmMakKevr02}.

When dealing with systems that would be described by (in our case, unavailable)
\emph{partial} differential equations, one can also reduce the {\it spatial} complexity.
For systems with one space dimension, the \emph{gap-tooth scheme}
\cite{KevrGearHymKevrRunTheo02} was proposed; it can be directly
generalized in several space dimensions.
A number of small intervals, separated by
large gaps, are introduced; they qualitatively correspond to mesh
points for a traditional, continuum solution of the unavailable
 equation. In higher space dimensions, these intervals would become
\emph{boxes} around the coarse mesh points, a term that we will also use
throughout this paper.
We construct a coarse time-$\Delta t$
map as follows. We first choose
a number of macroscopic grid points. Then, we choose a small
interval around each grid point; initialize the fine scale,
microscopic solver within each interval consistently with the
macroscopic initial condition profiles; and provide each box with
appropriate (as we will see, to some extent artificial) boundary
conditions.
Here, we constrain the macroscopic gradient to a value that is
determined by the macroscopic solution field.
Subsequently, we use the microscopic model in
each interval to simulate until time $\Delta t$, and
obtain macroscopic information (e.g.\ by computing the average
density in each box) at time $\Delta t$. This amounts to a coarse
time-$\Delta t$ map; this procedure is then repeated.

This ``coarse"  scheme has been used with
lattice-Boltzmann simulations of the Fitzhugh-Nagumo dynamics
\cite{Kevr00,KevrGearHymKevrRunTheo02} and with
particle-based simulations of the viscous Burgers equation
 \cite{GearLiKevr03}.
It was analyzed in the case where both the microscopic and the
macroscopic model are pure diffusion
\cite{KevrGearHymKevrRunTheo02}, where it was shown to be equivalent to
a standard finite difference scheme of order $2$ in space,
combined with an explicit Euler step in time.
Here, we extend the
analysis for the gap-tooth scheme in several ways. We derive a
formulation which approximates difference schemes that have higher
order accuracy in space; and we analyze the convergence of this
generalized scheme for a one-dimensional parabolic homogenization problem
with non-linear reaction. In this case, the microscopic model is a
partial differential equation with rapidly oscillating
coefficients. The macroscopic model is the \emph{effective}
equation that describes the evolution of the average behaviour.
In the limit, when the period of the oscillations becomes zero,
this effective equation is the classical homogenized equation.
The goal of the gap-tooth scheme is to approximate
the effective equation by using only the microscopic problem
inside the small boxes.  We analyze the accuracy of the method analytically
for the case where the homogenized solution is close to
the effective solution.
This analysis is important, because it
shows that the gap-tooth scheme approximates the correct
effective equation in the presence of microscopic scales.

It is worth mentioning that many numerical schemes have been
devised for the homogenization problem. Hou and Wu developed the
multi-scale finite element method that uses special basis functions to
capture the correct microscopic behaviour \cite{HouWu97,HouWu99}.
Schwab, Matache and Babuska have devised a generalized FEM method
based on a two-scale finite element space
\cite{SchwabMat02,MatBabSchwab00}.
Runborg et al.~\cite{RunTheoKevr02}
proposed a time-stepper based method that obtains the effective behaviour
through short bursts of detailed simulations appropriately averaged
over many shifted initial conditions.
The simulations were performed over the whole domain, but
the notion of effective behaviour is identical.
The guiding principle in equation-free timestepper-based computation
is to perform numerical tasks on an {\it unavailable} equation.
The time derivative for the evolution of the field is not obtained
from a formula, but estimated from observations of short,
appropriately initialized
and processed detailed dynamic simulations in (portions of) space.
When more information about the structure of the unavailable
equation is known (e.g. that it is a conservation law for a known observable),
it makes sense to modify the general time-stepper based procedure appropriately;
one can, for example, estimate the
time derivative based on flux computations using
an available microscopic simulator in (portions of) space.
This modification of equation-free computations for the case of
conservation laws forms the basis of the generalized Godunov scheme
of E and Enguist \cite{EEng03} and of the finite difference
heterogeneous multiscale method of Abdulle and E \cite{AbdE03}.
Our approach has focused on the general case where the structure
of the unavailable equation is not known.
It is interesting to pose the question about how one might know whether
a system is effectively a conservation law (and additional questions,
such as whether a system is Hamiltonian, or, possibly, integrable).
A computer-assisted methodology for the equation-free exploration of
such questions is introduced in \cite{LiKevrGearKevr03}.

In the gap-tooth scheme discussed here, the microscopic computations are
performed without assuming such a form for the ``right-hand-side'' of
the unavailable macroscopic equation; we evolve the detailed model
in a subset of the domain, and try to recover macroscopic
information by interpolation in space and extrapolation in time.
We note again that the gap-tooth scheme as it is presented here, is only
a part of the equation-free solution framework.
In this paper we examine the properties of this coarse time-stepper
{\it per se}; yet one should
keep in mind that the coarse time-$\Delta t$ map will
eventually be used inside a projective integration code, or
a bifurcation/continuation code.
The combination of gap-tooth timestepping with projective integration
has been termed  \emph{patch dynamics}
\cite{KevrGearHymKevrRunTheo02}.

 This paper is organized as follows.  In
section \ref{sec:gaptooth}, we discuss a general order formulation
of the gap-tooth scheme.  Subsequently, in section \ref{sec:3}, we
discuss some basic theoretical results on mathematical
homogenization, and we give a relation between the averaged
solution and the homogenized solution.  In section
\ref{sec:convergence}, we analyze the convergence of our method
for the model homogenization problem.
Numerical results
confirming the theorem are shown in section \ref{sec:numerical_results}.
This section also contains some examples for which the theory is
strictly speaking not valid.
We discuss a modified version of the gap-tooth scheme in section \ref{sec:dispersion} that avoids
constraining the macroscopic gradient during simulation. We introduce
so-called \emph{buffer} regions that shield
the dynamics inside each box from boundary effects.  At the outer boundary
of the buffer box, one can subsequently apply \emph{whatever boundary conditions
the microscopic code allows}.
We propose to study the resulting scheme by the (equation-free) numerical
computation of its damping factors.
%
%Integration with nearby coarse initial conditions is used to
%estimate matrix-vector products of the linearization of the coarse time-$\Delta t$ map
%with known perturbation vectors; these are integrated in matrix-free iterative
%methods such as Arnoldi eigensolvers to obtain its eigenvalues.
We show
how this can be done for a diffusion problem with Dirichlet boundary conditions.
We conclude in section
\ref{sec:conclusions}, where we also point out some next steps of
this research.

\section{The gap-tooth scheme\label{sec:gaptooth}}

We consider a general reaction-convection-diffusion equation with
a dependence on a small parameter $\epsilon$,
\begin{equation}\label{eq:general_homogenization_problem}
\frac{\partial}{\partial
t}u_{\epsilon}(x,t)=f\left(u_{\epsilon}(x,t),\frac{\partial}{\partial
x}u_{\epsilon}(x,t),\frac{\partial^2}{\partial
x^2}u_{\epsilon}(x,t),x,\frac{x}{\epsilon} \right),
\end{equation}
with initial condition $u_{\epsilon}(x,0)=u_0(x)$ and Dirichlet boundary
conditions $u_{\epsilon}(0,t)=v_o$ and $u_{\epsilon}(1,t)=v_1$.  We further assume that
$f$ is $1$-periodic in $y=\frac{x}{\epsilon}$.

We are only interested in the macroscopic (averaged)
behavior $u(x,t)$, which is a ``filtered'' version of $u_{\epsilon}(x,t)$.
To this end, we define an averaging operator for $u_{\epsilon}(x,t)$ as
follows,
\begin{equation}
U(x,t):=\mathcal{S}_h(u)(x,t)=\frac{1}{h}\int_{x-\frac{h}{2}}^{x+\frac{h}{2}}u_{\epsilon}(\xi,t)\d\xi.
\end{equation}
This operator replaces the unknown function by its local average
in a small box of size $h>>\epsilon$ around each point.  If $h$ is
sufficiently small, $U(x,t)$ should be a reasonable approximation to $u(x,t)$.

The averaged solution $U(x,t)$ satisfies an (unknown) evolution law, which we
assume also diffusive,
\begin{equation}\label{eq:general_macroscopic_equation}
\frac{\partial}{\partial
t}U(x,t)=F\left(U(x,t),\frac{\partial}{\partial
x}U(x,t),\frac{\partial^2}{\partial x^2}U(x,t),x;h \right).
\end{equation}
Note that this equation depends on the box width $h$.

The goal of the gap-tooth scheme is to approximate the solution
$U(x,t)$, while only making use of the detailed model
(\ref{eq:general_homogenization_problem}).  Suppose we want to
obtain the solution of the \emph{unknown}
 equation
\eqref{eq:general_macroscopic_equation} on the interval $[0,1]$,
using an equidistant macroscopic mesh $\Pi(\Delta
x):=\{0=x_0<x_1=x_0+\Delta x<\ldots<x_N=1\}$. We construct a time $\Delta t$-map
for $U(x,t)$ in the following way. We consider a small box
(\emph{tooth}) of length $h<<\Delta
x$ centered around each mesh point, and solve the original
problem \eqref{eq:general_homogenization_problem} in each box. To
determine the simulation within each box completely, we impose
 boundary constraints and an initial condition as follows.

\paragraph{Boundary constraints.}
Each box should provide information on the evolution of the global problem
at that location in space.
It is therefore crucial that the (artificially imposed)
boundary conditions
are chosen to emulate the correct behaviour in a larger domain.
Since the microscopic model
(\ref{eq:general_homogenization_problem}) is diffusive, it makes
sense (thinking of traditional explicit numerical schemes)
 to impose a fixed macroscopic concentration gradient at the
boundary of each small box during a time interval of length
$\Delta t$.
We determine the value of
this gradient by an approximation of the macroscopic
concentration profile $u(x,t)$ by a polynomial, based on the (given) box
averages $U_i^n$, $i=1,\ldots, N$.
\begin{displaymath}
u(x,t_n)\approx p_i^k(x,t_n),\qquad x \in
[x_i-\frac{h}{2},x_i+\frac{h}{2}],
\end{displaymath}
where $p_i^k(x,t_n)$ denotes a polynomial of (even) degree $k$. We
require that the approximating polynomial has the same box
averages as the initial condition in box $i$ and in $\frac{k}{2}$
boxes to the left and to the right.  This gives us
\begin{equation}\label{eq:polynomial_condition}
\frac{1}{h}\int_{x_{i+j}-\frac{h}{2}}^{x_{i+j}+\frac{h}{2}}
p_i^k(\xi,t_n)\d\xi=U_{i+j}^n, \qquad
j=-\frac{k}{2},\ldots,\frac{k}{2}.
\end{equation}
One can easily check that
\begin{equation}\label{eq:lagrange}
\mathcal{S}_{h}(p_i^k)(x,t_n)=\sum_{j=-\frac{k}{2}}^{\frac{k}{2}}U_{i+j}^n
L_{i,j}^k(x), \qquad
L_{i,j}^k(x)=\prod_{\stackrel{l=-\frac{k}{2}}{ l\ne
j}}^{\frac{k}{2}}\frac{(x-x_{i+l})}{(x_{i+j}-x_{i+l})}
\end{equation}
where $L_{i,j}^k(x)$ denotes a Lagrange polynomial of degree $k$.
The derivative of this approximating polynomial is subsequently
used to obtain the value of the gradient at the boundary of the
box.
\begin{equation}
\left.\frac{\partial p_i^k}{\partial x}\right|_{x_i\pm
\frac{h}{2}}=s_i^{\pm}
\end{equation}

If we did have an equation for the macroscopic behaviour, we would use these slopes as
Neumann boundary conditions.  Here, we use these derivatives to constrain the
\emph{average} gradient of the
detailed solution $u(x,t)$ in box $i$ over one small-scale period
around the end points,
\begin{equation}\label{eq:bc}
\frac{1}{\epsilon}\int_{x_i-\frac{h}{2}-\frac{\epsilon}{2}}^{x_i-\frac{h}{2}+\frac{\epsilon}{2}}\frac{\partial}{\partial
\xi}u(\xi,t)\d\xi =s_i^{-}, \qquad
\frac{1}{\epsilon}\int_{x_i+\frac{h}{2}-\frac{\epsilon}{2}}^{x_i+\frac{h}{2}+\frac{\epsilon}{2}}\frac{\partial}{\partial
\xi}u(\xi,t)\d\xi =s_i^{+}.
\end{equation}

Note that we approximate a box average in a box of
macroscopic size $h >> \epsilon$, while we average for boundary
condition purposes
over a length scale $\epsilon$ that is characteristic for the
microscopic model.  Hence, we replace each boundary
condition and its effect on the simulation by an algebraic constraint.

\paragraph{Initial conditions.}
For the time integration, we must impose an initial condition
$\tilde{u}^i(x,t_n)$ in each box
$[x_i-\frac{h}{2},x_i+\frac{h}{2}]$, at time $t_n$. We require
$\tilde{u}^i(x,t_n)$ to satisfy the boundary conditions and the
given box average.  We choose a quadratic polynomial $\tilde{u}(x,t_n)$, centered
around the coarse mesh point $x_i$,
\begin{equation}\label{eq:lifting_quad}
\tilde{u}^i(x,t_n)\equiv a(x-x_i)^2+b(x-x_i)+c.
\end{equation}
Using the constraints \eqref{eq:bc} in the limit for $\epsilon \to 0$ and requiring
\begin{displaymath}
\frac{1}{h}\int_{x_{i}-\frac{h}{2}}^{x_i+\frac{h}{2}}\tilde{u}^i(\xi,t_n)\d\xi
=U_i^n,
\end{displaymath}
 we obtain
\begin{equation}
a=\frac{s_i^+ -s_i^-}{2h},\qquad b=\frac{s_i^+ +s_i^-}{2}, \qquad
c={U}_i^n-\frac{h}{24}(s_i^+ - s_i^-)\label{eq:lifting}.
\end{equation}

\begin{figure}
\begin{center}
\includegraphics[scale=0.6]{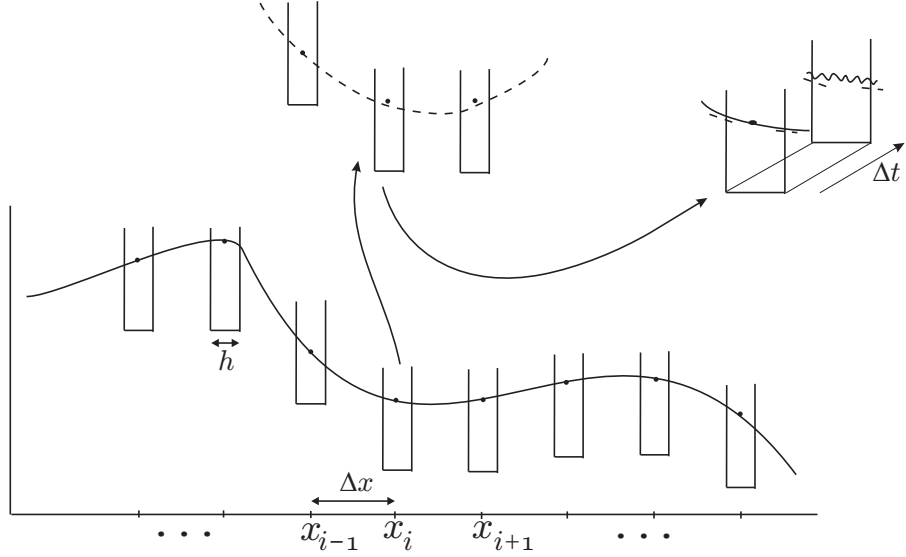}
\caption{\label{fig:schematic} A schematic representation of
a gap-tooth time-step.  We choose a number of boxes of size $h$
around each macroscopic mesh point $x_i$ and interpolate the
initial averages (dots) in a number of boxes around $x_i$ (dashed
profile). The derivatives at the boundary and the average are used
to create an initial profile in box $i$ (full line).}
\end{center}
\end{figure}

\paragraph{The algorithm.}  The complete \emph{gap-tooth} algorithm
to proceed from $U^n$ to $U^{n+1}$ is given below (see also figure
\ref{fig:schematic}):
\begin{enumerate}
\item \textbf{Lifting.}  At time $t_n$, construct
the initial condition $\tilde{u}^i(x,t_n)$, $i=0,\ldots,N$, using
the box averages $U^n_j$ ($j=0,\ldots, N$) as defined in
(\ref{eq:lifting}).
\item \textbf{Evolution.}  Compute $\tilde{u}^i(x,t)$ by solving the equation
(\ref{eq:general_homogenization_problem}) until time
$t_{n+1}=t+\Delta t$ with the boundary constraints (\ref{eq:bc}).
\item \textbf{Restriction.}  Compute the box average
$U_i^{n+1}=\frac{1}{h}\int_{x_i-\frac{h}{2}}^{x_i+\frac{h}{2}}
\tilde{u}_{\epsilon}(\xi,t_{n+1})\d\xi$ at time $t_{n+1}$.
\end{enumerate}
It is clear that this amounts to a ``coarse-to-coarse'' time-$\Delta
t$ map. We write this map as follows,
\begin{equation}
U^{n+1}=S_k(U^n;t_n+\Delta t),
\end{equation}
where $S$ represents the numerical time-stepping scheme for the
macroscopic (coarse) variables and $k$ denotes the degree of
interpolation.

\paragraph{Microscopic simulators.}  It is possible that the microscopic
model is not a partial differential equation, but some microscopic simulator,
e.g.~kinetic Monte Carlo or molecular dynamics code.  In fact, this is the
case where our method would be most useful.  In this case, several complications
arise.  First of all, the choice of the box width $h$ becomes important, since
there will generally exist a trade-off between statistical accuracy (e.g.~enough sampled
particles) and spatial resolution.

Second, the \emph{lifting} step, i.e.~the construction
of box initial conditions, also becomes more involved.  In general, the microscopic
model will have many more degrees of freedom, the \emph{higher order moments} of the
evolving distribution.  These will quickly become slaved to the governing moments (the
ones where the lifting is conditioned upon), see e.g.~\cite{KevrGearHymKevrRunTheo02,MakMarKevr02},
but it might be better to do a constrained run before initialization
to create mature initial conditions \cite{HumKevr02,Constraint03}.

Third, as already mentioned, imposing macroscopically inspired boundary
conditions is non-trivial \cite{LiYip98}.
Moreover a given microscopic code
may come with one of several \emph{``standard"} microscopic boundary
conditions.
We will therefore examine the effect of incorporating simulations
with such standard
boundary conditions in a gap-tooth context, provided we extend the simulation
in a buffer region surrounding the computational ``tooth".
The solution in the buffer is not used in the restriction step.
This variant is examined more closely in section \ref{sec:dispersion}.

Finally, even determining which and how many macroscopically
inspired boundary conditions
are needed, is a delicate issue.
This is related with
the order of the partial differential
equation, i.e.~the order of the highest spatial derivative.
A systematic way
to estimate this, without having the macroscopic equation,
is given in \cite{LiKevrGearKevr03}.

\section{Model homogenization problem}\label{sec:3}

Here, we review some basic results from homogenization theory.  We note
that we are interested in finding the \emph{effective} behaviour of
the solution.  In our setup, we know that for sufficiently small $\epsilon$
the effective behaviour is close to the \emph{homogenized} behaviour,
which is the limit of the solution for $\epsilon\to 0$.
Since in some cases, the homogenized equation can be found analytically,
we use this equation as our reference for the effective behaviour.

\subsection{Standard homogenization theory\label{sec:homogenization}}
As a model problem, we consider the following parabolic partial
differential equation,
\begin{equation}\label{eq:model_equation}
\frac{\partial}{\partial
t}u_{\epsilon}(x,t)=\frac{\partial}{\partial
x}\left(a\left(\frac{x}{\epsilon}\right)\frac{\partial}{\partial
x}u_{\epsilon}(x,t) \right)+g(u_\epsilon(x,t)),
\end{equation}
with initial condition $u_{\epsilon}(x,0)=u^0(x)$ and suitable
boundary conditions. In this equation,
$a(y)=a\left(\frac{x}{\epsilon}\right)$ is periodic in $y$ and
$\epsilon$ is a small parameter.

Consider equation \eqref{eq:model_equation} with Dirichlet
boundary conditions $u_{\epsilon}(0,t)=v_0$ and
$u_{\epsilon}(1,t)=v_1$. According to classical homogenization theory
\cite{BenLioPap78}, the solution to
\eqref{eq:model_equation} can be written as an asymptotic
expansion in $\epsilon$,
\begin{equation}\label{eq:asymptotic_expansion}
u_{\epsilon}(x,t)=u_0(x,t)+\sum_{i=1}^{\infty}\epsilon^i \left(
u_i(x,\frac{x}{\epsilon},t)\right),
\end{equation}
where the functions $u_i(x,y,t)\equiv
u_i(x,\frac{x}{\epsilon},t)$, $i=1,2,\ldots$ are periodic in $y$.
Here, $u_0(x,t)$ is the solution of the \emph{homogenized
equation}
\begin{equation}\label{eq:hom_eq}
\frac{\partial}{\partial t}u_0(x,t)=\frac{\partial}{\partial
x}\left(a^*\frac{\partial}{\partial x}u_0(x,t)\right)+g(u_0(x,t))
\end{equation}
with initial condition $u_0(x,0)=u^0(x)$ and Dirichlet boundary
conditions $u_0(0,t)=v_0$ and $u_0(1,t)=v_1$; $a^*$ is the
constant effective coefficient, given by
\begin{equation}
a^*=\int_0^1 a(y)\left(1-\frac{\d}{\d y}\chi(y)\right)\d y,
\end{equation}
and $\chi(y)$ is the periodic solution of
\begin{equation}\label{eq:cell_problem}
\frac{\d}{\d y}\left(a(y)\frac{\d}{\d
y}\chi(y)\right)=\frac{\d}{\d y}a(y),
\end{equation}
the so-called \emph{cell problem}.  The solution of
\eqref{eq:cell_problem} is only defined up to an additive
constant, so we impose the extra condition
\begin{equation}
\int_0^1 \chi(y)\d y=0.
\end{equation}
From this cell problem, we can derive $u_1(x,y,t)=\frac{\partial
u_0}{\partial x}\chi(y)$.

These asymptotic expansions have been rigorously justified in the
classical book \cite{BenLioPap78}.  Under appropriate smoothness
assumptions, one can obtain pointwise convergence of $u_0$ to
$u_{\epsilon}$ as $\epsilon \to 0$.  Therefore, we can write
\begin{equation}\label{eq:pointwise_convergence}
\left\|u_{\epsilon}(x,t)-u_0(x,t)\right\|\le C_0\epsilon,
\end{equation}
where
$\left\|f(x)\right\|\equiv\left\|f(x)\right\|_{\infty}=\max_{x}|f(x)|$
denotes the $\infty$-norm of $f$.  Throughout this text, whenever
we use $\left\|\cdot\right\|$, we mean the $\infty$-norm.

It is important to note that the gradient of $u(x,t)$ is given by
\begin{equation}
\frac{\partial}{\partial
x}u_{\epsilon}(x,t)=\frac{\partial}{\partial
x}u_0(x,t)+\frac{\partial}{\partial y}u_1(x,y,t)+O(\epsilon),
\end{equation}
from which it is clear that the micro-scale fluctuations have a
strong effect on the local detailed gradient. Nevertheless, since $u_i(x,y,t)$
are periodic in $y$, the gradient of the homogenized solution can
be approximated by the averaged gradient over one period
$\epsilon$ of the medium.  The error is bounded by
\begin{equation}
\left\| \frac{\partial}{\partial x}u_0(x,t)-
\frac{1}{\epsilon}\int_{x-\frac{\epsilon}{2}}^{x+\frac{\epsilon}{2}}
\frac{\partial}{\partial x}u_{\epsilon}(\xi,t)\d\xi\right\| \le
C_0' \epsilon.
\end{equation}

\subsection{Homogenization and averaging\label{sec:averaging}}

The gap-tooth scheme introduces an approximation on two levels.
The scheme computes an approximation to the evolution of the
averaged macroscopic quantities instead of an approximation to the
solution of the true homogenized solution. Before considering how
well the gap-tooth scheme approximates this averaged behaviour, it
might be of interest to show how the averaged behaviour
approximates the homogenized solution.

\begin{lem}\label{lem:asymptotic_expansion_of_averages}
For $u(x,t)$ sufficiently smooth, the averaged function
\begin{displaymath}
U(x,t)=\mathcal{S}_h(u)(x,t):=\frac{1}{h}\int_{x-\frac{h}{2}}^{x+\frac{h}{2}}u(\xi,t)\d\xi
\end{displaymath}
can be asymptotically expanded in $h$ as follows,
\begin{displaymath}
U(x,t)=u(x,t)+\sum_{l=1}^{\infty}\left(\frac{h}{2}\right)^{2l}\frac{1}{(2l+1)!}
\left.\frac{\partial^{2l}}{\partial^{2l}
\xi}u(\xi,t)\right|_{\xi=x}
\end{displaymath}
\end{lem}
We omit the proof, but this can easily be checked using Maple.

Using this lemma, we consider the homogenization problem of
section \ref{sec:homogenization},
\begin{equation}
\frac{\partial}{\partial
t}u_{\epsilon}(x,t)=\frac{\partial}{\partial
x}\left(a\left(\frac{x}{\epsilon}\right)\frac{\partial}{\partial
x}u_{\epsilon}(x,t) \right)+g(u_{\epsilon}(x,t))
\end{equation}
In this case, we can bound the difference between the averaged
solution $U(x,t)$ and the homogenized solution $u_0(x,t)$ in the
following way.
\begin{lem}\label{lem:averaging_error}
The difference between the homogenized solution $u_0(x,t)$ and the
averaged solution
$U(x,t)=\int_{x-\frac{h}{2}}^{x+\frac{h}{2}}u(\xi,t)\d\xi$ is
bounded by
\begin{displaymath}
\left\|U(x,t)-u_0(x,t)\right\|\le C_1 h^2+C_2\epsilon.
\end{displaymath}
\end{lem}
\proof We first make use of the asymptotic expansion
\eqref{eq:asymptotic_expansion} for $u(x,t)$ and the triangle
inequality, and subsequently of lemma
\ref{lem:asymptotic_expansion_of_averages}.
\begin{eqnarray*}
\left\|U(x,t)-u_0(x,t)\right\|&=&
\left\|\frac{1}{h}\int_{x-\frac{h}{2}}^{x+\frac{h}{2}}u(\xi,t)\d\xi-u_0(x,t)\right\|\\
&\le&\left\|\frac{1}{h}\int_{x-\frac{h}{2}}^{x+\frac{h}{2}}u_0(\xi,t)
\d\xi-u_0(x,t)\right\|\\&&+\epsilon\left\|
\int_{x-\frac{h}{2}}^{x+\frac{h}{2}}u_1(\xi,\frac{\xi}{\epsilon},t)-
\theta_1(\xi,\frac{\xi}{\epsilon},t)\d\xi\right\|+O(\epsilon^2)\\
 &\le&\frac{h^2}{24}\left|\frac{\partial^2}{\partial
 x^2}u_0(x,t)\right|+C_2\epsilon\\
 &\le&\frac{h^2}{24}\max_{x\in[0,1]}\left|\frac{\partial^2}{\partial
 x^2}u_0(x,t)\right|+C_2\epsilon
\end{eqnarray*}
This concludes the proof.
\endproof  This shows
that the averaged solution is a good approximation of the
homogenized solution for sufficiently small box width $h$.

\section{Convergence results \label{sec:convergence}}

To analyze the convergence of the gap-tooth scheme, we solve the
detailed problem approximately in each box.  Because
$h>>\epsilon$, we can resort to the homogenized solution, and
bound the error using equation \eqref{eq:pointwise_convergence}.
It is important to note that we only use the homogenized equation
for analysis purposes. We never make use of
the homogenized equation in the implementation.

We first relate the gap-tooth time-stepper as constructed in section
\ref{sec:gaptooth} with a gap-tooth time-stepper for which
the box problem is the homogenized equation with Neumann boundary
conditions.

\begin{lem}\label{lem:equivalent_box_problem}
Consider the model equation, %\eqref{eq:model_equation},
\begin{equation}\label{eq:model_eq_repeated}
\frac{\partial}{\partial
t}u_{\epsilon}(x,t)=\frac{\partial}{\partial
x}\left(a\left(\frac{x}{\epsilon}\right)\frac{\partial}{\partial
x}u_{\epsilon}(x,t) \right)+g\left(u_{\epsilon}(x,t)\right),
\end{equation}
where $a(y)=a\left(\frac{x}{\epsilon}\right)$ is periodic in $y$
and $\epsilon$ is a small parameter, with initial condition
$u_{\epsilon}(x,0)=u^0(x)$ and boundary constraints
\begin{equation}\label{eq:bc_r}
\frac{1}{\epsilon}\int_{x_i-\frac{h}{2}-\frac{\epsilon}{2}}^{x_i-\frac{h}{2}+\frac{\epsilon}{2}}\frac{\partial}{\partial
\xi}u_{\epsilon}(\xi,t)\d\xi =s_i^{-}, \qquad
\frac{1}{\epsilon}\int_{x_i+\frac{h}{2}-\frac{\epsilon}{2}}^{x_i+\frac{h}{2}+\frac{\epsilon}{2}}\frac{\partial}{\partial
\xi}u_{\epsilon}(\xi,t)\d\xi =s_i^{+}.
\end{equation}
For $\epsilon\to 0$, this problem converges to the homogenized
problem
\begin{equation}\label{eq:h_r}
\frac{\partial}{\partial t}u_0(x,t)=\frac{\partial}{\partial
x}\left(a^*\frac{\partial}{\partial x}u_0(x,t)\right)+g(u_0(x,t))
\end{equation}
with initial condition $u_0(x,0)=u^0(x)$ and Neumann boundary
conditions
\begin{equation}\label{eq:h_bc_r}
\left.\frac{\partial}{\partial x
}u_0(x,t)\right|_{x=x_{i}\pm\frac{h}{2}}=s_i^{\pm},
\end{equation}
and the solution of
\eqref{eq:model_eq_repeated}-\eqref{eq:bc_r} converges pointwise
to the solution of \eqref{eq:h_r}-\eqref{eq:h_bc_r}, with the
following error estimate
\begin{equation}
\left\|u_{\epsilon}(x,t)-u_0(x,t)\right\|\le C_3\epsilon.
\end{equation}
\end{lem}
This lemma can be checked using the two-scale convergence method
\cite{All92} or formally by making use of asymptotic expansions
\cite{CioDon99}.

Using this lemma, we now estimate the difference between a
gap-tooth time-step using
(\ref{eq:model_eq_repeated}-\ref{eq:bc_r}) and a gap-tooth
time-step using the homogenized box problem
(\ref{eq:h_r}-\ref{eq:h_bc_r}).

\begin{lem}
Define $U^{n+1}=S_k(U^{n},t_n+\Delta t)$ as one gap-tooth
time-step on the full problem \eqref{eq:model_eq_repeated} with
the box constraints \eqref{eq:bc_r}, and
$\hat{U}^{n+1}=\hat{S}_k(\hat{U}^n,t_n+\Delta t)$ as one gap-tooth
time-step on the homogenized problem \eqref{eq:h_r} with boundary
conditions \eqref{eq:h_bc_r}. When $U^n=\hat{U}^n$, we have
\begin{displaymath}
\left\|U_i^{n+1}-\hat U_i^{n+1}\right\|\le C_4\epsilon.
\end{displaymath}
\end{lem} \proof
Denote the solutions of
\eqref{eq:model_eq_repeated}-\eqref{eq:bc_r} and
\eqref{eq:h_r}-\eqref{eq:h_bc_r}, with the initial condition
$\tilde{u}_i(x,t_n)$ determined by the lifting step,
\eqref{eq:lifting_quad}-\eqref{eq:lifting} as
$\tilde{u}_{\epsilon}^i(x,t)$ resp.\ $\tilde{u}_0^i(x,t)$. We can
then write
\begin{eqnarray*}
\left\|U_i^{n+1}-\hat{U}_i^{n+1}\right\|&=&
\left\|\frac{1}{h}\int_{x_i-\frac{h}{2}}^{x_i+\frac{h}{2}}
\tilde{u}_{\epsilon}^i(\xi,t_{n+1})\d \xi -
\frac{1}{h}\int_{x_i-\frac{h}{2}}^{x_i+\frac{h}{2}}\tilde{u}_0^i(\xi,t_{n+1})\d
\xi \right\| \\
&=&\frac{1}{h}\left\|\int_{x_i-\frac{h}{2}}^{x_i+\frac{h}{2}}
\left(\tilde{u}_{\epsilon}(\xi,t_{n+1})-\tilde{u}_0^i(\xi,t_{n+1})\right)\d \xi\right\|\\
&\le& \max_{x \in [x_i-\frac{h}{2},x_i+\frac{h}{2}]}
\left\|\tilde{u}_{\epsilon}(\xi,t_{n+1})-\tilde{u}_0^i(\xi,t_{n+1})\right\| \\
&\le&C_3 \epsilon
\end{eqnarray*}
Here, we bounded the average over the interval
$[x_i-\frac{h}{2},x_i+\frac{h}{2}]$ by the maximum, and
subsequently used lemma \ref{lem:equivalent_box_problem}. This is
valid since we assumed $U^n=\hat{U}^n$. Therefore, the initial
condition for both box problems is the same.

  This proves the
lemma.
\endproof

The averaged solution $U(x,t)$ satisfies a reaction-diffusion-like equation
\begin{equation}\label{eq:avg_eq}
\frac{\partial}{\partial x}U(x,t)=\frac{\partial ^2}{\partial x^2}U(x,t)+\frac{1}{h}\int_{x-\frac{h}{2}}^{x+\frac{h}{2}}g(u(\xi,t))\d\xi.
\end{equation}
We denote a forward Euler/spatial finite difference approximation for (\ref{eq:avg_eq})  as
\begin{displaymath}
\bar{U}^{n+1}=\bar{S}_k(\bar{U}^n,t_n+\Delta t),
\end{displaymath}
with $k$ the order of accuracy of the spatial finite differences.
The following theorem compares
a gap-tooth time-step $\hat{U}^{n+1}=\hat{S}_k(\hat{U}^n,t_n+\Delta t)$ with a
finite difference time-step.

\begin{lem}\label{thm:gaptooth_timestep}
We denote a finite difference approximation of order $k$ for the
evolution of $U(x,t)$ as
\begin{displaymath}
\bar{U}^{n+1}=\bar{S}_k(\bar{U}^n,t_n+\Delta t),
\end{displaymath}
and one gap-tooth time-step with homogenized box problems (\ref{eq:h_r}-\ref{eq:h_bc_r})
as $\hat{U}^{n+1}=\hat{S}_k(\hat{U}^n,t_n+\Delta t)$.  The exact solution of the homogenized
equation is denoted by $u(x,t)$.  When
$\hat{U}^n=\bar{U}^n=\mathcal{S}_h(u)(x,t_n)$, we have the following estimate
\begin{displaymath}
\left\|\hat{U}^{n+1}_i-\bar{U}^{n+1}_i\right\| \le C_4 \Delta t^2.
\end{displaymath}
\end{lem}
\proof Denote the solution \eqref{eq:h_r}-\eqref{eq:h_bc_r} with
the initial condition $\tilde{u}_i(x,t_n)=a(x-x_i)^2+b(x-x_i)+c$,
determined by the lifting step
\eqref{eq:lifting_quad}-\eqref{eq:lifting}, as
$\tilde{u}_0^i(x,t)$.
\begin{itemize}
\item We write an expansion for the solution of
\eqref{eq:h_r}-\eqref{eq:h_bc_r} using the method of separation of
variables.  The solution can be decomposed as follows
\begin{displaymath}
\tilde{u}_0^i(x,t)=\tilde{u}_0^i(x,t_n)+\tilde{v}^i(x,t),
\end{displaymath}
where $\tilde{v}^i(x,t)$ is the solution of
\begin{displaymath}
\frac{\partial}{\partial
t}\tilde{v}^i(x,t)=\frac{\partial}{\partial
x}\left(a^*\frac{\partial}{\partial
x}\tilde{v}^i(x,t)\right)+a^*\cdot 2a+g(\tilde{u}_0^i(x,t)),
\end{displaymath}
with homogeneous initial condition and homogeneous Neumann
boundary conditions. The last term is due to the spatial
derivatives of the initial profile. We write $\tilde{v}^i(x,t)$ as
a Fourier series with time-dependent coefficients,
\begin{displaymath}
\tilde{v}^i(x,t)=\frac{a_0^i(t)}{2}+\sum_{n=1}^{\infty}a_n^i(t)\cos\left(k_n(x-x_i)\right)+
\sum_{n=0}^{\infty}b_n^i(t)\sin\left(l_n(x-x_i)\right),
\end{displaymath}
with
%\begin{displaymath}
$k_n=\frac{2n\pi}{h}$ %\qquad \text{
and
%} \qquad
$l_n=\frac{(2n+1)\pi}{h}$.
%\end{displaymath}
The Fourier modes satisfy the homogeneous Neumann boundary
conditions, and form a set of spatial basis functions for the
solution. The time-dependent coefficients are given by
\begin{eqnarray*}
a_n^i(t)&=&\frac{2}{h}\int_{x_i-\frac{h}{2}}^{x_i+\frac{h}{2}}\tilde{v}^i(\xi,t)\cos\left(\frac{2\pi
n}{h}(\xi-x_i)\right)\d\xi\\
b_n^i(t)&=&\frac{2}{h}\int_{x_i-\frac{h}{2}}^{x_i+\frac{h}{2}}\tilde{v}^i(\xi,t)\sin\left(\frac{(2n+1)\pi}{h}(\xi-x_i)\right)\d\xi\\
\end{eqnarray*}
Each coefficient can be found by solving an ordinary differential
equation that is obtained by taking the time derivative and
replacing the time derivative of the solution by the right-hand side
of the partial
differential equation.
\item We use this analytical solution
to obtain an explicit formula for one time-step of the
gap-tooth scheme.  We write
\begin{eqnarray}
\hat{U}_i^{n+1}&=&\frac{1}{h}\int_{x_i-\frac{h}{2}}^{x_i+\frac{h}{2}}\tilde{u}_0^i(\xi,t_n+\Delta t)\d\xi\notag\\
&=&\frac{1}{h}\int_{x_i-\frac{h}{2}}^{x_i+\frac{h}{2}} \left(
\tilde{u}_0^i(\xi,t_n)+\tilde{v}^i(\xi,t_n+\Delta t)\right)\d\xi\notag\\
&=&\hat{U}_i^n+\frac{a_0^i(t_n+\Delta t)}{2},
\end{eqnarray}
where the last step is due to the definition of
$\tilde{u}_0^i(x,t_n)$ and the zero averages of sine and cosine.
Thus, we only need to consider the coefficient $a_0^i(t)$ in what
follows.
\item
For $a_0(t)$, we get the following ordinary differential equation
\begin{displaymath}
\frac{\d}{\d
t}a_0^i(t)=\frac{2}{h}\int_{x_i-\frac{h}{2}}^{x_i+\frac{h}{2}}\frac{\partial}{\partial
t}\tilde{v}^i(\xi,t)\d\xi,
\end{displaymath}
with initial condition $a_0^i(t_n)=0$.  This yields
\begin{eqnarray*}
\frac{\d}{\d
t}a_0^i(t)&=&\frac{2}{h}\int_{x_i-\frac{h}{2}}^{x_i+\frac{h}{2}}\frac{\partial}{\partial
t}\tilde{v}^i(\xi,t)\d\xi\\
&=&\frac{2}{h}\int_{x_i-\frac{h}{2}}^{x_i+\frac{h}{2}}\frac{\partial}{\partial
\xi}\left(a^*\frac{\partial}{\partial
\xi}\tilde{v}^i(\xi,t)\right)+a^*\cdot 2a+g(\tilde{u}_{0}^i(\xi,t))\d\xi\\
&=&\frac{2}{h}\left(\left.a^*\frac{\partial}{\partial
\xi}\tilde{v}^i(\xi,t)\right|_{\xi=x_i-\frac{h}{2}}^{\xi=x_i+\frac{h}{2}}
+a^*\cdot 2ah
+\int_{x_i-\frac{h}{2}}^{x_i+\frac{h}{2}}g(\tilde{u}_{0}^i(\xi,t))\d\xi\right)\\
&=&4a^*\cdot a
+\frac{2}{h}\int_{x_i-\frac{h}{2}}^{x_i+\frac{h}{2}}g(\tilde{u}_{0}^i(\xi,t))\d\xi,
\end{eqnarray*}
where we could discard the first term because of the boundary
conditions.

The resulting formula for a gap-tooth time-step is
\begin{equation}\label{eq:gaptooth_timestep}
\hat{U}_i^{n+1}=\hat{U}_i^n+2a^*\cdot a\Delta t
+\frac{1}{h}\int_{t=t_n}^{t=t_n+\Delta t}
\int_{x_i-\frac{h}{2}}^{x_i+\frac{h}{2}}g(\tilde{u}_{0}^i(\xi,t))\d\xi
\d t
\end{equation}
\item
We now wish to connect one time-step of the gap-tooth scheme with
one finite difference time-step for the equation for the averaged
solution $U(x,t)$.  We first notice that
\begin{eqnarray*}
a&=&\frac{s_i^+-s_i^-}{2h}\\
&=&\frac{1}{2h}\left(\left.\frac{\partial p_k^i(x,t)}{\partial
x}\right|_{x=x_i+\frac{h}{2}}-\left.\frac{\partial
p_k^i(x,t)}{\partial
x}\right|_{x=x_i-\frac{h}{2}}\right)\\
&=\frac{1}{2}&\left.\frac{\partial^2}{\partial
\xi^2}\mathcal{S}_h(p_k^i)(x,t)\right|_{x=x_i}\\
&=&\frac{1}{2}\sum_{j=-\frac{k}{2}}^{\frac{k}{2}}{U_{i+j}^n}\left.\frac{\d^2}{\d
x^2} L_{i,j}^k(x)\right|_{x=x_i}
\end{eqnarray*}
Therefore, the second derivative of the
$k$-th order polynomial that interpolates the box averages is equal to $2a$.
Due to symmetry, the first and second derivatives of this polynomial in
$x_i$ are the standard finite difference approximation of order
$k$ of the function $\bar{U}(x,t_n)$.  This can easily be verified
with Maple.  This leads to the following formula for one finite
difference time-step of the equation for the averages
\begin{equation}\label{eq:finitedifference_timestep}
\bar{U}^{n+1}_i=\bar{U}_i^n+2a^*\cdot a \Delta
t+G_i^n\Delta t,
\end{equation}
where
\begin{displaymath}
G^n_i=\frac{1}{h}\int_{x_i-\frac{h}{2}}^{x_i+\frac{h}{2}}
g(u(\xi,t))\d\xi,
\end{displaymath}
with $u(x,t)$ the exact solution of (\ref{eq:hom_eq}).
\item
We now estimate
\begin{eqnarray*}
\lefteqn{\left\|\bar{U}^{n+1}_i-\hat{U}_i^{n+1}\right\|}\\
&=&\left\|G_i^n\Delta
t-\frac{1}{h}\int_{t_n}^{t_n+\Delta t}
\int_{x_i-\frac{h}{2}}^{x_i+\frac{h}{2}}g(\tilde{u}_{0}^i(\xi,t))\d\xi
\d t\right\|\\
&\le&\left\|\frac{1}{h}\int_{x_i-\frac{h}{2}}^{x_i+\frac{h}{2}}\left(
g\left(u(\xi,t_n)\right)\Delta t -\int_{t_n}^{t_n+\Delta t}
g\left(u(\xi,t)\right)\d t\right)\d\xi\right\|\\
&+&\left\|\frac{1}{h}\int_{x_i-\frac{h}{2}}^{x_i+\frac{h}{2}}\int_{t_n}^{t_n+\Delta t}
\left(g\left(u(\xi,t)\right)-g\left(\tilde{u}_0^i(\xi,t)\right)\right)\d t\d\xi\right\|
\end{eqnarray*}
The first term can be written as
\begin{eqnarray*}
\lefteqn{\left\|\frac{1}{h}\int_{x_i-\frac{h}{2}}^{x_i+\frac{h}{2}}\left(
g\left(u(\xi,t_n)\right)\Delta t -\int_{t_n}^{t_n+\Delta t}
g\left(u(\xi,t)\right)\d t\right)\d\xi\right\|}\\
&=&\Delta
t\left\|\frac{1}{h}\int_{x_i-\frac{h}{2}}^{x_i+\frac{h}{2}}\left(g(u(\xi,t_n))
-\frac{1}{\Delta t}\int_{t_n}^{t_n+\Delta
t}g(u(\xi,t))\d t\right)\d\xi\right\|\\
&\le&\Delta
t\left\|\frac{1}{h}\int_{x_i-\frac{h}{2}}^{x_i+\frac{h}{2}}\frac{1}{2}\left.\frac{\partial
g}{\partial u}\right|_{u=u(\xi,t_n)}\left.\frac{\partial
u}{\partial t}\right|_{t=t_n}\Delta
t\d\xi\right\|\\ &\le& C'\Delta t^2,
\end{eqnarray*}
where we used a Taylor expansion for $g$ and the chain rule.  The second term is estimated as
\begin{eqnarray*}
\lefteqn{\left\|\frac{1}{h}\int_{x_i-\frac{h}{2}}^{x_i+\frac{h}{2}}\int_{t_n}^{t_n+\Delta t}
\left(g\left(u(\xi,t)\right)-g\left(\tilde{u}_0^i(\xi,t)\right)\right)\d t\d\xi\right\|}\\
&\le&\left\|\frac{1}{h}\int_{x_i-\frac{h}{2}}^{x_i+\frac{h}{2}}\int_{t_n}^{t_n+\Delta t}
L\left(u(\xi,t)-\tilde{u}_0^i(\xi,t)\right)\d t\d\xi\right\|\\
&\le&L\left\|\frac{1}{h}\int_{x_i-\frac{h}{2}}^{x_i+\frac{h}{2}}
\int_{t_n}^{t_n+\Delta t}
\left(u(\xi,t_n)-\tilde{u}_i^0(\xi,t_n)\right)+(t-t_n)C\d t \d\xi\right\|\\
&\le&C''\Delta t^2,
\end{eqnarray*}
where we used Lipschitz continuity of $g$ and the fact that (due to lifting)
\begin{displaymath}
\int_{x_i-\frac{h}{2}}^{x_i+\frac{h}{2}}u(\xi,t_n)\d \xi=
\int_{x_i-\frac{h}{2}}^{x_i+\frac{h}{2}}\tilde{u}_0^i(\xi,t_n)\d \xi.
\end{displaymath}
This proves the lemma.
\end{itemize}
\endproof

We now have the following result.
\begin{thm}[Local error]\label{thm:truncation_error}
Define $U^{n+1}=S_k(U^n,t_n+\Delta t)$ as the result of one
gap-tooth time-step for (\ref{eq:model_equation}), and
$\bar{U}^{n+1}=\bar{S}_k(\bar{U}^n,t_n+\Delta t)$ as a finite difference time-step
for (\ref{eq:avg_eq}).  When
$U^n_i=\bar{U}_i^n=\mathcal{S}_h(u)(x_i,t_n)$ (the exact solution
at $(x_i,t_n)$ ), the difference is bounded by
\begin{displaymath}
\left\|U_i^{n+1}-\bar{U}_i^{n+1}\right\|\le C_3\epsilon+C_9\Delta
t^2
\end{displaymath}
\end{thm}
\begin{proof}
This follows immediately by combining lemmas \ref{lem:equivalent_box_problem}
and \ref{thm:gaptooth_timestep}.
\end{proof}

Therefore, we obtain the following error bound.

\begin{thm}
If $\bar{U}^{n+1}=\bar{S}_k(\bar{U}^n,t_n+\Delta t)$ is a stable finite difference
scheme for (\ref{eq:avg_eq}), then $U^{n+1}=S_k(U^n,t_n+\Delta t)$.  Moreover,
if $U^0=\bar{U}^0=\mathcal{S}_h(u^0)(x)$, the error with respect to the homogenized solution $u_0(x,t)$ of
(\ref{eq:hom_eq}) is bounded by
\begin{displaymath}
\left\|U_i^{n}-u_0(x_i,t_n)\right\|\le C_1 h^2 + C_2 \epsilon + C_8 \frac{\epsilon}{\Delta t} + C_9\Delta t + C_5 \Delta x^k
\end{displaymath}
\end{thm}
\begin{proof}
We start by splitting the error, based on the origin of the error contributions,
\begin{eqnarray*}
\lefteqn{\left\|U_i^{n}-u_0(x_i,t_n)\right\|}\\
&\le&\left\|U_i^n-\bar{U}_i^n\right\|+\left\|\bar{U}_i^n-U(x_i,t_n)\right\|+\left\|U(x_i,t_n)+u_0(x_i,t_n)\right\|\\
&\le&\left\|U_i^n-\bar{U}_i^n\right\|+C_5\Delta x^k+C_6\Delta t+C_1 h^2 + C_2 \epsilon.
\end{eqnarray*}
The last two terms follow from standard finite difference theory and from lemma \ref{lem:asymptotic_expansion_of_averages}.
The first term merits further investigation,
\begin{eqnarray*}
\left\|U^n-\bar{U}^n\right\|&=&\left\|S_k(U^{n-1},t_{n-1}+\Delta t)-\bar{S}_k(\bar{U}^{n-1},t_{n-1}+\Delta t)\right\|\\
&\le&\left\|S_k(U^{n-1},t_{n-1}+\Delta t)-\bar{S}_k(U^{n-1},t_{n-1}+\Delta t)\right\|\\
&+&\left\|\bar{S}_k(U^{n-1},t_{n-1}+\Delta t)-\bar{S}_k(\bar{U}^{n-1},t_{n-1}+\Delta t)\right\|\\
&\le&\left\|A_k(U^{n-1}-\bar{U}^{n-1})\right\|+C_3\epsilon+C_4\Delta t^2,
\end{eqnarray*}
where the last line is due to lemma \ref{thm:truncation_error} and $A_k$ is the error amplification matrix
for the forward Euler/spatial finite difference scheme.  Therefore, stability of the finite difference scheme is
necessary and sufficient
to bound the errors from previous steps. By induction, we obtain
\begin{eqnarray*}
\left\|U^n-\bar{U}^n\right\|&\le& C n (C_3\epsilon+C_4\Delta t^2)\\
&\le& C \frac{t_n}{\Delta t}(C_3\epsilon+C_4\Delta t^2)
\end{eqnarray*}
We prove the theorem by combining all terms.
\end{proof}

The result clearly shows the interplay between the different approximations; we have an error
due to the (macroscopic) finite difference scheme, an error that arises because we consider
box averages, and an extra error due to the setup of the box problems in each time-step.
The introduction of averaged Neumann boundary conditions generates an error which is
independent of the time-step.  We therefore
have to make a trade-off between the accuracy that is gained by taking shorter time-steps
and the accuracy that is lost because of more frequent reinitializations.  This will also be
shown in the numerical experiments.  Projective integration \cite{GearKevr01} can
help in reducing this error, since then less re-initializations are needed.

\section{Numerical results\label{sec:numerical_results}}

We show convergence of the gap-tooth method for a diffusion
problem with a rapidly oscillating diffusion coefficient (section
\ref{sec:5.1}), a reaction-diffusion system (section
\ref{sec:5.2}), and a system with a rough non-periodic (random) diffusion
coefficient (section \ref{sec:5.3}).

\subsection{Periodic diffusion coefficient without reaction \label{sec:5.1}}

\begin{figure}
\begin{center}
\subfigure{\includegraphics[scale=0.8]{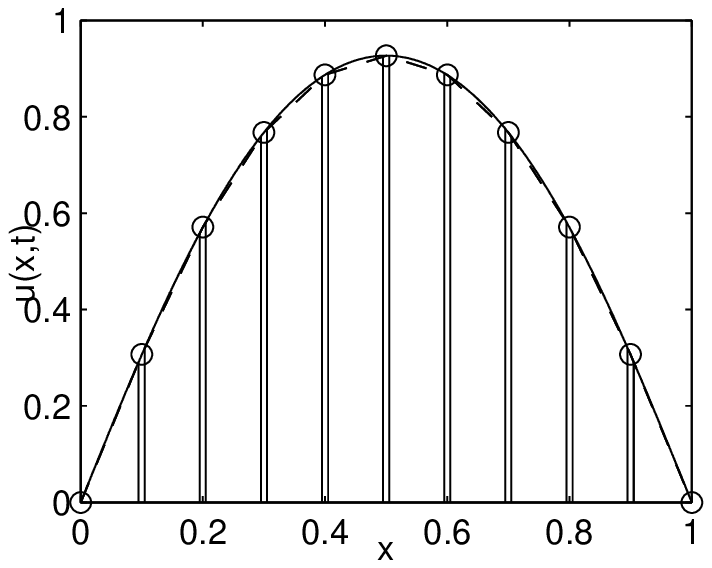}}
\subfigure{\includegraphics[scale=0.8]{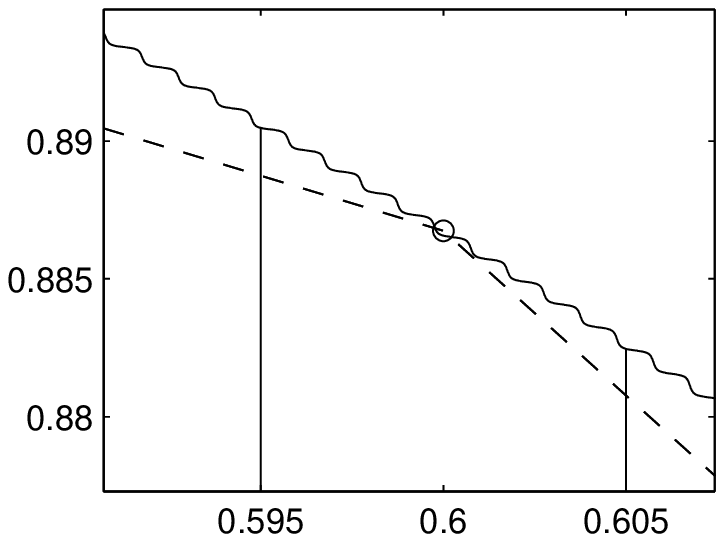}}
\caption{\label{fig:solution_diff}Left: Solution of equation
(\ref{eq:ex1_fine}) with an implicit Euler scheme with $\delta
x=1\cdot 10^{-5}$ and $\delta t=5\cdot 10^{-7}$ (full line), and a
gap-tooth solution (dashed) with $\Delta x=0.05$, $\Delta
t=2.5\cdot 10^{-4}$ and $h=0.01$, with internal finite differences
as for the full problem. The boxes indicate the domains where the
computations are done in the gap-tooth scheme. Right: A zoom shows
the presence of the microscopic fluctuations.}
\end{center}
\end{figure}

Consider the following model problem,
\begin{equation}\label{eq:ex1_fine}
\frac{\partial}{\partial
t}u_{\epsilon}(x,t)=\frac{\partial}{\partial
x}\left(a(\frac{x}{\epsilon})\frac{\partial}{\partial
x}u_{\epsilon}(x,t) \right),\qquad
a(\frac{x}{\epsilon})=1.1+\sin(2\pi \frac{x}{\epsilon})
\end{equation}
with $\epsilon=1\cdot 10^{-3}$, $x\in [0,1]$, initial conditions
$u_{\epsilon}(x,0)=1-4(x-0.5)^2$, and Dirichlet boundary
conditions $u_{\epsilon}(0,t)=u_{\epsilon}(1,t)=0$.  To solve the
microscopic problem, we use a standard finite difference
discretization in space and an implicit Euler time-step, with
parameters $\delta x=1\cdot 10^{-5}$ and $\delta t=5\cdot
10^{-7}$. The corresponding homogenized equation is given by
\begin{equation}\label{eq:ex1_coarse}
\frac{\partial}{\partial x}\left(a^*\frac{\partial}{\partial
x}u_0(x,t)\right), \qquad a^*\approx 0.45825686.
\end{equation}
With respect to the theoretical setup of section
\ref{sec:convergence}, two additional approximations are made
during the computations: the time integration inside each box is
not exact; and we have to use numerical quadrature formulas to
obtain the box average at each restriction.  The resolution
of the internal time-stepper is such that these effects are
negligible with respect to the other sources of error that we
wish to study.
In our code, we used the
trapezoidal rule as quadrature formula.
Figure \ref{fig:solution_diff} shows the
solution of (\ref{eq:ex1_fine}) and the gap-tooth solution with
$\Delta x=0.1$, $\Delta t=1\cdot 10^{-3}$ and $h=0.01$ at time
$t=0.02$.

\paragraph{Difference with respect to finite differences.}
We first compare the results of the gap-tooth scheme for
(\ref{eq:ex1_fine}) with those of a finite difference scheme for
(\ref{eq:ex1_coarse}) with the same coarse parameters. We use a
spatial interpolation/finite differences of order $k=2$, with a
coarse mesh of $\Delta x=0.1$, resp. $\Delta x=0.05$, for
decreasing box sizes $h=0.04$, $0.02$, $0.01$, $0.005$, and for
time-steps $\Delta t=\nu \Delta x^2$, with $\nu=0.1$, $0.2$,
$0.4$.  The results are shown in table \ref{tab:diff-fd-gt-1} for
$\Delta x=0.1$ and in table \ref{tab:diff-fd-gt-2} for $\Delta
x=0.05$.  They clearly show an $O(h^2)$ decrease of the error
initially, with a slow-down for smaller $h$, due to the additional
$O(\frac{\epsilon}{\Delta t})$ term.  We also see that the decrease of convergence
speed is affected by the time-step.  The error decreases less
rapidly for smaller $\nu$, due to the additional error in each
restriction step. Also note a smaller decrease for $\Delta x=0.05$, because
in this case $\Delta t$ is also smaller.
Note that the difference with respect to finite
differences does not depend dramatically on $\Delta t$ for this example, due to
the absence of a reaction term (see theorem
\ref{thm:truncation_error}).  By comparing tables
\ref{tab:diff-fd-gt-1} and \ref{tab:diff-fd-gt-2}, we also see
that the difference between the gap-tooth scheme and the finite
difference scheme is independent of $\Delta x$, for fixed $h$ and $\nu$.
\begin{table}
\begin{center}
\begin{tabular}{|l||r|r||r|r||r|r|} \hline
   & \multicolumn{2}{|c||}{$\Delta t=0.1 \Delta x^2$}
   & \multicolumn{2}{|c||}{$\Delta t=0.2 \Delta x^2$} &
   \multicolumn{2}{|c|}{$\Delta t=0.4 \Delta x^2$}\\
   \hline
   & error & ratio & error & ratio & error & ratio \\ \hline
  $h=0.04$ & $5.4189 \cdot 10^{-4}$ &       & $5.3755\cdot 10^{-4}$ &     &   $5.3568\cdot
  10^{-4}$ &
  \\\hline
  $h=0.02$ & $1.4296\cdot 10^{-4}$ &       3.79  & $1.3815\cdot 10^{-4}$ &  3.89    &  $1.3584\cdot
  10^{-4}$ & 3.94
  \\\hline
  $h=0.01$ & $4.3169\cdot 10^{-5}$ &   3.32       & $3.8297\cdot 10^{-5}$ &  3.61    & $3.5885\cdot
  10^{-5}$ &  3.79
  \\\hline
  $h=0.005$ & $1.8221\cdot 10^{-5}$ &  2.37       & $1.3334\cdot 10^{-5}$ &   2.87 &   $1.0896\cdot
  10^{-5}$ & 3.29\\
  \hline
\end{tabular}
\caption{Difference between the gap-tooth scheme for
(\ref{eq:ex1_fine}) and a finite difference scheme for
(\ref{eq:ex1_coarse}) for order $k=2$ and $\Delta x=0.1$ at time
$t=2\cdot 10^{-2}$.\label{tab:diff-fd-gt-1}}
\end{center}
\end{table}
\begin{table}
\begin{center}
\begin{tabular}{|l||r|r||r|r||r|r|} \hline
   & \multicolumn{2}{|c||}{$\Delta t=0.1 \Delta x^2$}
   & \multicolumn{2}{|c||}{$\Delta t=0.2 \Delta x^2$} &
   \multicolumn{2}{|c|}{$\Delta t=0.4 \Delta x^2$}\\
   \hline
   & error & ratio & error & ratio & error & ratio \\ \hline
  $h=0.04$ & $5.6378 \cdot 10^{-4}$ &       & $5.5060\cdot 10^{-4}$ &     &   $5.4275\cdot
  10^{-4}$ &
  \\\hline
  $h=0.02$ & $1.7152\cdot 10^{-4}$ &   3.29      & $1.5293\cdot 10^{-4}$ &  3.6    &  $1.5641\cdot
  10^{-4}$ & 3.79
  \\\hline
  $h=0.01$ & $7.2618\cdot 10^{-5}$ &    2.36      & $5.3027\cdot 10^{-5}$ & 2.88     & $4.3236\cdot
  10^{-5}$ &  3.32
  \\\hline
  $h=0.005$ & $4.7638\cdot 10^{-5}$ &   1.52      & $2.8043\cdot 10^{-5}$ & 1.89   &   $1.8247\cdot
  10^{-5}$ & 2.37 \\
  \hline
\end{tabular}
\caption{Difference between the gap-tooth scheme for
(\ref{eq:ex1_fine}) and a finite difference scheme for
(\ref{eq:ex1_coarse}) for order $k=2$ and $\Delta x=0.05$ at time
$t=2\cdot 10^{-2}$.\label{tab:diff-fd-gt-2}}
\end{center}
\end{table}

\begin{figure}
\begin{center}
\includegraphics[scale=0.5]{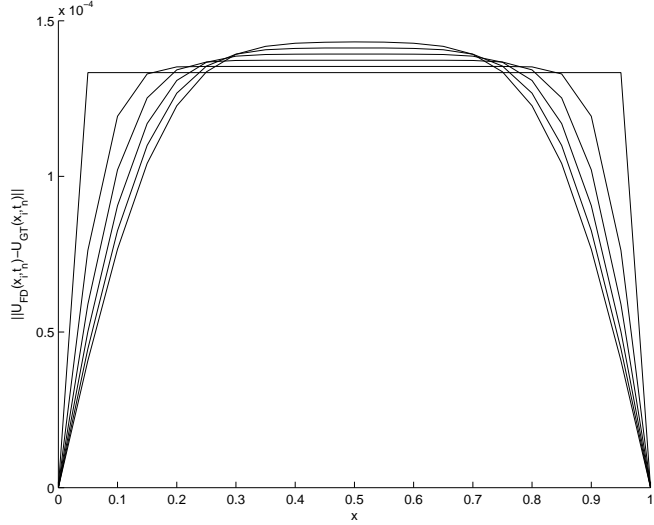}
\caption{Difference between the gap-tooth scheme for
(\ref{eq:ex1_fine}) with $k=2$, $\Delta x=0.05$, $\Delta t=1\cdot
10^{-3}$, $h=0.02$, and a finite difference scheme with the same
coarse parameters for time $t=0$, $4\cdot 10^{-3}$, $8\cdot
10^{-3}$, $12\cdot 10^{-3}$, $16 \cdot 10^{-3}$, $2\cdot
10^{-2}$\label{fig:error_gaptooth_k2}.}
\end{center}
\end{figure}

\paragraph{The $O(\frac{\epsilon}{\Delta t})$-term.}
The evolution of the difference between the gap-tooth scheme and
the finite difference scheme is shown in figure
\ref{fig:error_gaptooth_k2}.
We see that we start with a constant error at time $t=0$ due to
the averaging of the initial condition. Note that this
error is not important if one compares to the exact
\emph{averaged} solution $U(x,t)$.  It is an artifact of comparing to
the homogenized equation instead of the effective equation.
However, if $u_0(x,t)$
evolves according to (\ref{eq:ex1_coarse}),
$U(x,t)=\mathcal{S}_h(u_0)(x,t)$ evolves according to the same
equation.  Therefore, we can eliminate the $O(h^2)$ term, by
comparing to $U(x,t)$ instead of $u_0(x,t)$.  This allows us to
show that the stagnation in tables \ref{tab:diff-fd-gt-1} and
\ref{tab:diff-fd-gt-2} is really $\epsilon$-dependent.
 We compare the results of the gap-tooth
scheme and the finite difference scheme of order $2$ for $\Delta x=0.05$ and
$h=0.02$ at time $t=2\cdot 10^{-2}$.  We first keep $\Delta t=1\cdot 10^{-3}$
fixed, and vary $\epsilon=1\cdot 10^{-3}$, $2\cdot 10^{-3}$, $4\cdot 10^{-3}$.
Subsequently, we fix $\epsilon=1\cdot 10^{-3}$, and vary $\Delta t=0.5\cdot 10^{-3}$,
$1\cdot 10^{-3}$, $2\cdot 10^{-3}$.  The results are shown in table \ref{tab:eps_over_deltat}.
For this simple example, the error decreases
quadratically with $\epsilon$, because the error constant of the
$O(\epsilon)$ term is zero.  If we combine the results from both tables, we clearly see a
decrease in error according to
$O(\frac{\epsilon^2}{\Delta t})$ for this example.
\begin{table}
\begin{tabular}{c c}
\begin{tabular}{|l||r|r|}
\hline & error \phantom{blah}& ratio \\
\hline $\epsilon=4\cdot 10^{-3}$ & $3.0574\cdot 10^{-4}$ &  \\
$\epsilon=2\cdot 10^{-3}$ & $7.6239\cdot 10^{-5}$ & 4.01\\
$\epsilon=1\cdot 10^{-3}$ & $1.9710\cdot 10^{-5}$ & 3.87 \\
\hline
\end{tabular}
&
\begin{tabular}{|l||r|r|}
\hline & error \phantom{blah}& ratio \\
\hline $\Delta t=0.5\cdot 10^{-3}$ & $3.9304\cdot 10^{-5}$ &  \\
$\Delta t=1\cdot 10^{-3}$ & $1.9710\cdot 10^{-5}$ & 1.99\\
$\Delta t=2\cdot 10^{-3}$ & $9.9133\cdot 10^{-6}$ & 1.99 \\
\hline
\end{tabular}
\end{tabular}
\caption{
Left: Difference between
the results of the gap-tooth scheme and the forward Euler/spatial finite difference
scheme of order $2$ for $\Delta x=0.05$, $\Delta t=1\cdot
10^{-3}$, $h=0.02$, at time $t=2\cdot 10^{-2}$, for
$\epsilon=1\cdot 10^{-3}$, $2\cdot 10^{-3}$, $4\cdot 10^{-3}$
after subtracting the $O(h^2)$ error in the initial data. \label{tab:eps_over_deltat}
Right: Difference for $\epsilon=1\cdot 10^{-3}$, $\Delta t$ varying.}
\end{table}

\begin{figure}
\begin{center}
\subfigure[$k=2$, $\Delta
x=0.1$]{\includegraphics[scale=0.7]{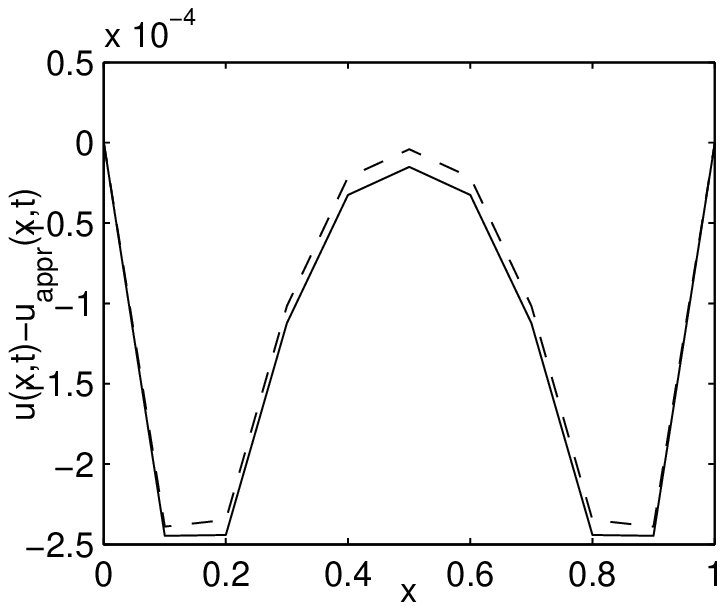}}\subfigure[$k=2$,
$\Delta
x=0.05$]{\includegraphics[scale=0.7]{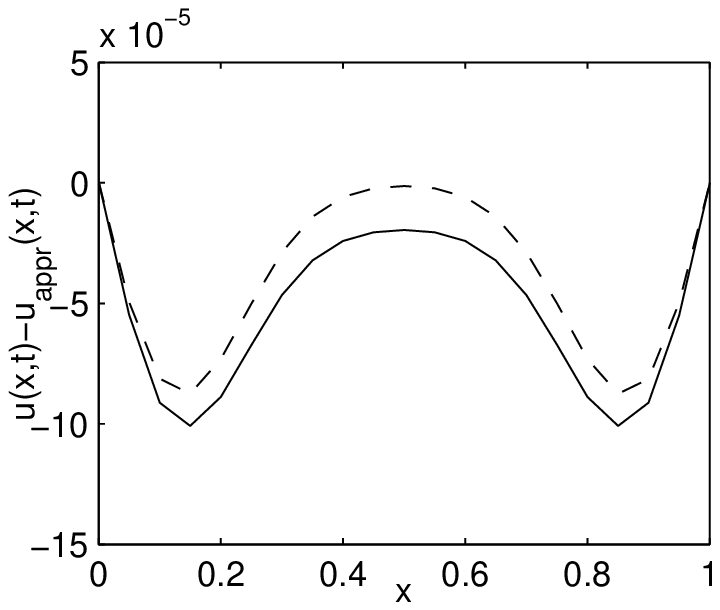}}
\caption{\label{fig:diff-ex-gt}Error of gap-tooth (full line) with
$h=0.005$ and finite differences (dashed) with respect to an exact solution for the homogenized
equation for $\Delta x=0.1$ (left), $\Delta x=0.05$ (right) and
$\Delta t=0.4\Delta x^2$.}
\end{center}
\end{figure}

\paragraph{Error with respect to solution of the homogenized problem.}
We also show the error with respect to the exact solution of the
homogenized problem. For
this purpose, we compute the homogenized solution with a second
order finite difference approximation in space with $\delta
x=1\cdot 10^{-5}$ and implicit Euler time-steps with $\delta
t=5\cdot 10^{-7}$.  The gap-tooth scheme is used with box width
$h=0.005$, $\Delta x=0.2$, $0.1$ and $\Delta t=\nu\Delta x^k$ with
$\nu=0.4$ and order $k=2$.  We compare the gap-tooth
and the finite difference solution to the exact solution for the homogenized
equation at time $t=2\cdot 10^{-2}$.
The results are shown in
figure \ref{fig:diff-ex-gt}.  It is clear that the error is
similar to that of the finite difference scheme.  Note however
that the errors will increase when the $O(h^2)$ and
$O({\epsilon}{\Delta t})$ terms in the error become dominant.

\begin{figure}
\subfigure{\includegraphics[scale=0.35]{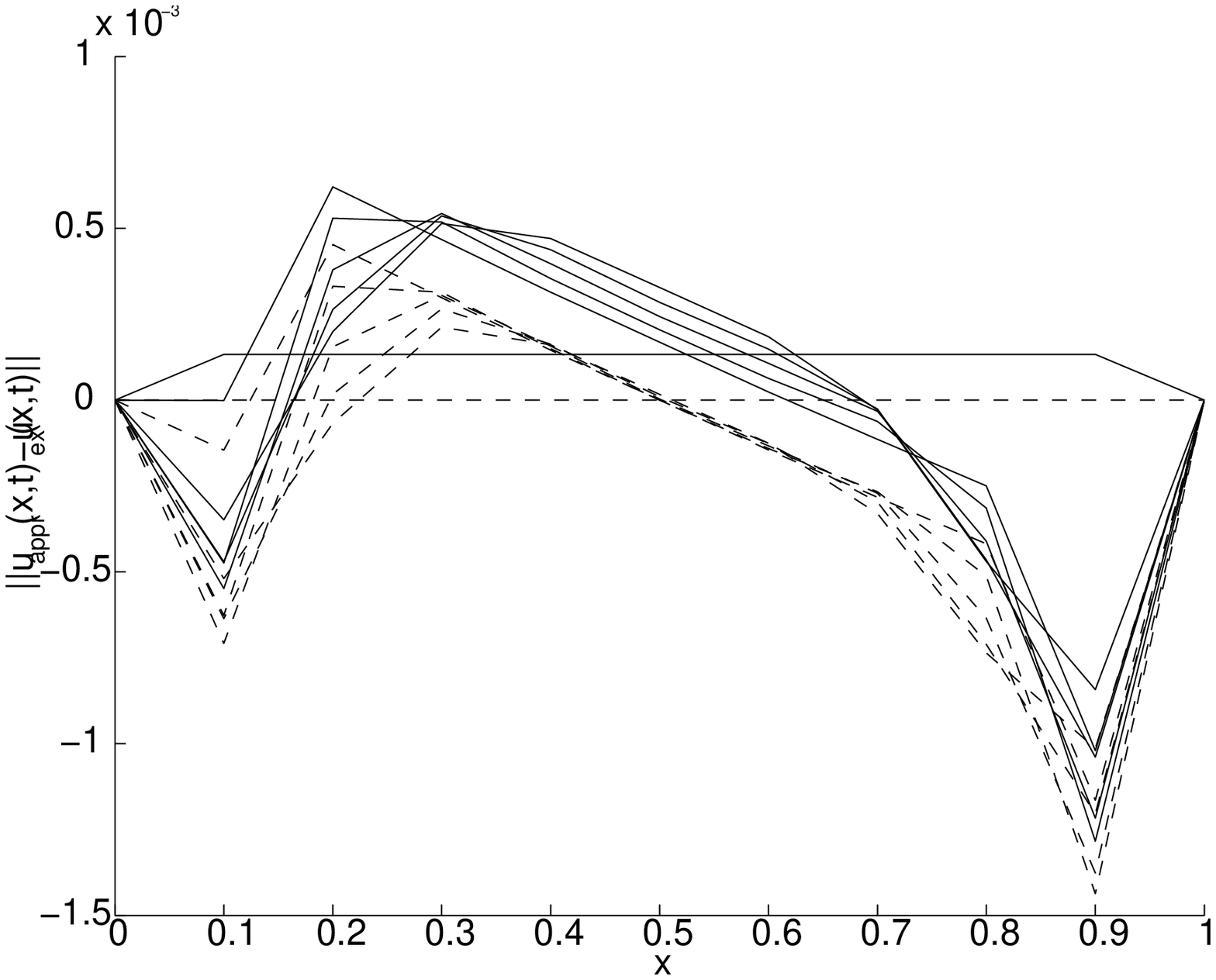}}
\subfigure{\includegraphics[scale=0.35]{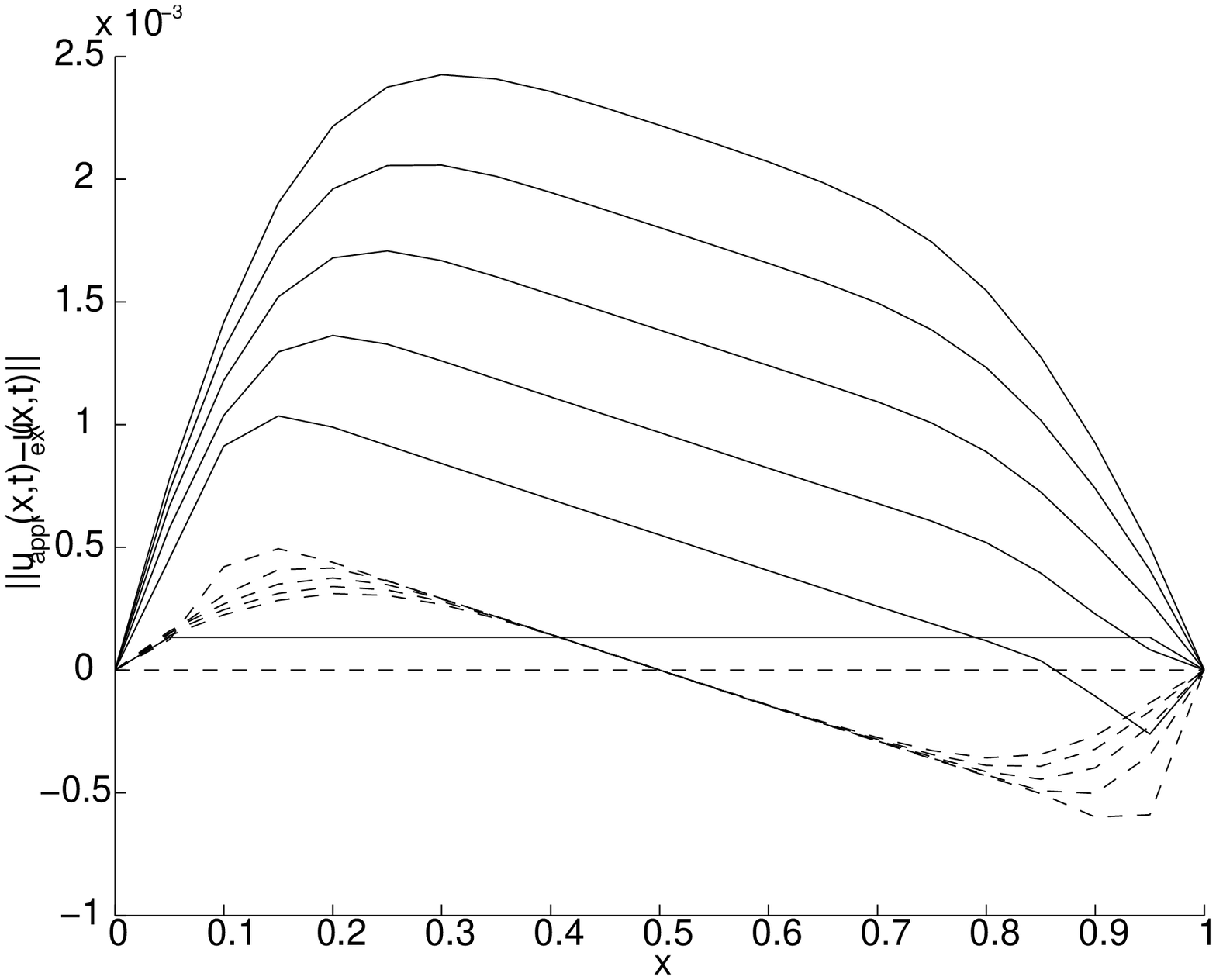}}
\caption{\label{fig:fourth_order}Error of the gap-tooth scheme (full line) and a
finite difference scheme, both of order 4, with respect to the exact homogenized solution at times
$t=4\cdot 10^{-3}$, $8\cdot 10^{-3}$, $12\cdot 10^{-3}$, $16\cdot 10^{-3}$, $2\cdot 10^{-2}$,
where $\Delta x=0.1$ (left), $\Delta x=0.05$ (right), $\Delta t=0.4\Delta x^4$ and $h=0.01$.}
\end{figure}

\paragraph{Higher order discretizations in space.}
We repeat the same experiment for a gap-tooth scheme of order
$k=4$, which we compare to a fourth order spatial finite
difference approximation, with an explicit Euler time-step.  As parameters,
we choose $\Delta x=0.1$, $0.05$ and $\Delta
t=\nu \Delta x^4$, with $\nu=0.4$ and $h=0.01$.
In order to view the $O(\Delta x^4)$ behaviour, we need to choose $\Delta t$
correspondingly small, which will affect convergence through the
$O(\frac{\epsilon}{\Delta t})$ term.  The results are shown in figure \ref{fig:fourth_order}.
It is clear for $\Delta x=0.1$ that the scheme approximates the fourth order scheme.  In this case,
the time-step $\Delta t= 4\cdot 10^{-5}$.  However,
for $\Delta x=0.05$, $\Delta t=2.5\cdot 10^{-6}$, the error is already completely dominated by
the $O(\frac{\epsilon}{\Delta t})$ term.

\subsection{Periodic diffusion with non-linear reaction \label{sec:5.2}}
\begin{figure}
\begin{center}
\includegraphics[scale=0.3]{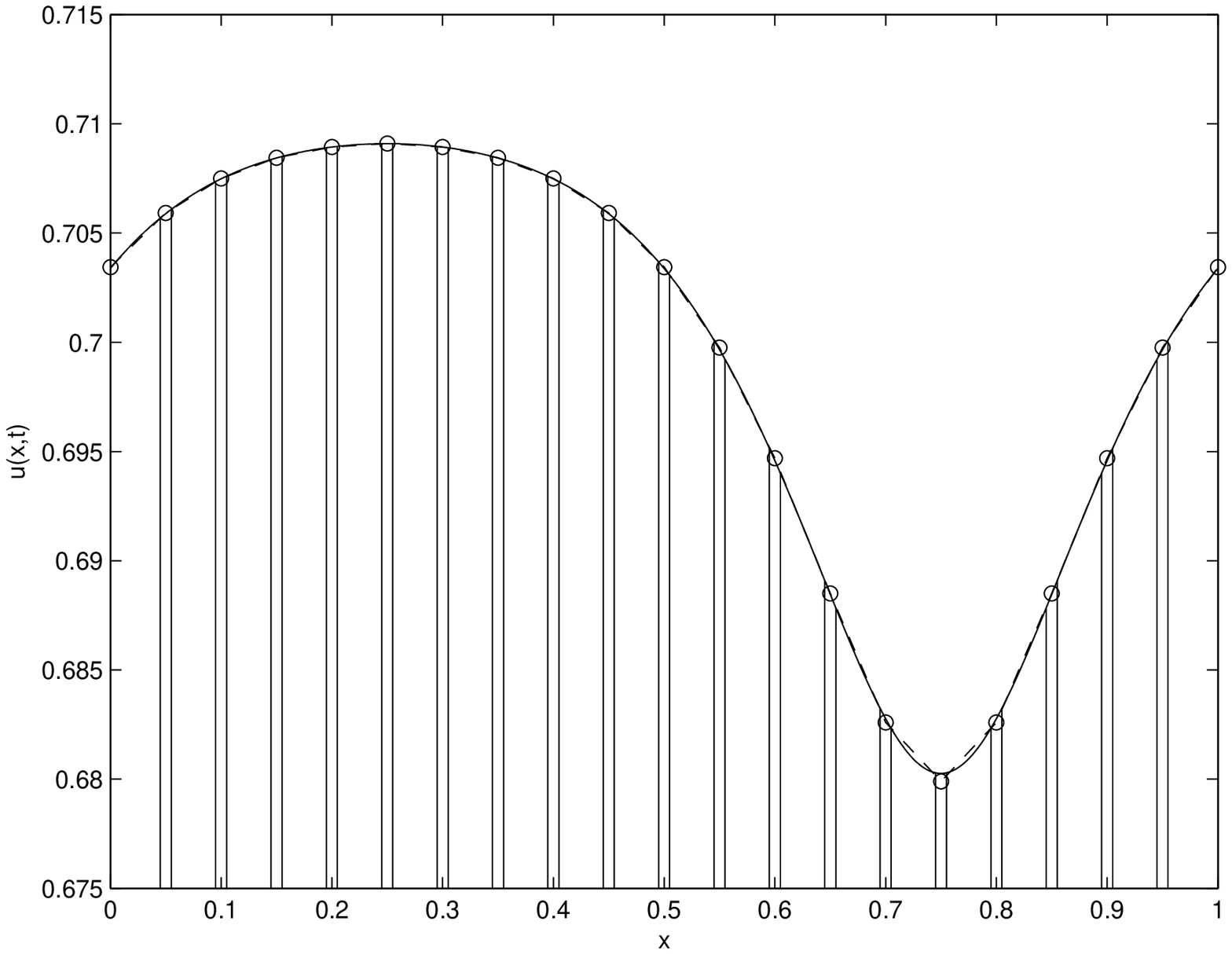}
\includegraphics[scale=0.3]{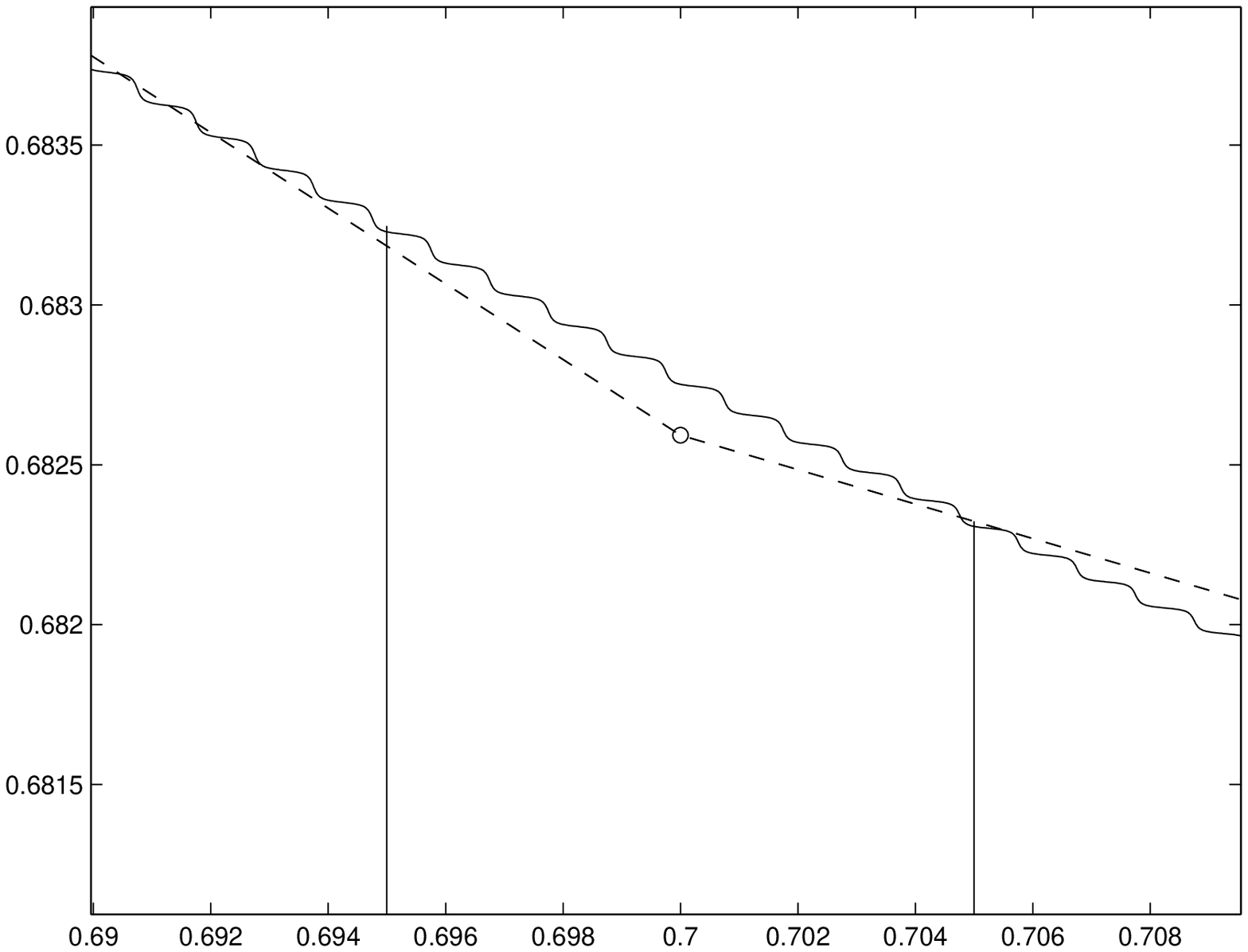}
\caption{\label{fig:reac_diff} Reference solution (full line) and gap-tooth solution (dashed/circles)
for the problem (\ref{eq:ex2_fine}) at time $t=2\cdot 10^{-2}$.}
\end{center}
\end{figure}

As a second example, we consider the following reaction-diffusion equation
\begin{equation}\label{eq:ex2_fine}
\frac{\partial}{\partial t}u_{\epsilon}(x,t)=\frac{\partial}{\partial x}\left(
a(\frac{x}{\epsilon})\frac{\partial}{\partial}u_{\epsilon}(x,t)\right)+
u_{\epsilon}(x,t)(1-\frac{u_{\epsilon}(x,t)}{b(x)}),
\end{equation}
where $a(\frac{x}{\epsilon})=1.1+\sin(2\pi\frac{x}{\epsilon})$ as in section \ref{sec:5.1}, and
$b(x)=\sin(2\pi x)+1.2$.  This model can be interpreted as a one-species logistic growth
model with macroscopically varying capacity $b(x)$ and a rapidly oscillating diffusion coefficient
$a(\frac{x}{\epsilon})$.  We choose periodic boundary conditions and a constant initial condition,
$u^0(x)=0.7$.
The corresponding homogenized problem is given by
\begin{equation}\label{eq:ex2_coarse}
\frac{\partial}{\partial t}u_{\epsilon}(x,t)=\frac{\partial}{\partial x}\left(
a^*\frac{\partial}{\partial}u_{\epsilon}(x,t)\right)+
u_{\epsilon}(x,t)(1-\frac{u_{\epsilon}(x,t)}{b(x)}),
\end{equation}
with  $a^*\approx 0.45825686$.
Figure \ref{fig:reac_diff} shows the solution of (\ref{eq:ex2_fine}), as well as the result of
a gap-tooth simulation with parameters $\Delta x=0.05$, $\Delta t=1\cdot 10^{-3}$, $h=0.01$ for
$t=2\cdot 10^{-2}$.
The reference solution was computed with second order spatial finite differences and an implicit
Euler time-stepper, with $\delta x=1\cdot 10^{-5}$ and $\delta t=5\cdot 10^{-7}$.

In contrast to the example in section \ref{sec:5.1}, there will be also an $O(\Delta t)$ difference
between the gap-tooth solution and the finite difference scheme.  To show this, we compare the gap-tooth
scheme with $\Delta x=0.1$ and $\Delta t=\nu \Delta x^2$ with a finite difference scheme
for the homogenized equation with the same coarse parameters for $\nu=0.05$, $0.1$, $0.2$, $0.4$.  We
did this for box width $h=0.005$ and $0.01$.  The results are shown in table \ref{tab:diff-fd-gt-1-reac}.
\begin{table}
\begin{center}
\begin{tabular}{|l||r|r||r|r|} \hline
   & \multicolumn{2}{|c||}{$h=0.005$}
   & \multicolumn{2}{|c|}{$h=0.01$}\\
   \hline
   &  error \phantom{blah} & ratio & error \phantom{blah} & ratio \\ \hline
  $\Delta t=4\cdot 10^{-3}$ & $1.3842\cdot 10^{-4}$ &     &   $1.3929\cdot  10^{-4}$ &
  \\\hline
  $\Delta t=2\cdot 10^{-3}$   & $7.9135\cdot 10^{-5}$ &  1.75    &  $7.9792\cdot  10^{-5}$ & 1.75
  \\\hline
  $\Delta t=1\cdot 10^{-3}$  & $5.1103\cdot 10^{-5}$ & 1.55     & $5.1496\cdot   10^{-5}$ &  1.55
  \\\hline
  $\Delta t=5\cdot 10^{-4}$   & $3.8014\cdot 10^{-5}$ & 1.34   &   $3.7959\cdot  10^{-5}$ & 1.36 \\
  \hline
\end{tabular}
\caption{Difference between the gap-tooth scheme for
(\ref{eq:ex1_fine}) and a finite difference scheme for
(\ref{eq:ex1_coarse}) for order $k=2$ and $\Delta x=0.05$ at time
$t=2\cdot 10^{-2}$.\label{tab:diff-fd-gt-1-reac}}
\end{center}
\end{table}
>From this table we see that a smaller time-step indeed gives a smaller difference with respect to
the corresponding finite difference scheme.  However, we do not observe the ratio $2$.
We can show that this is due to the interference
of the $O(\frac{\epsilon}{\Delta t})$-term.  Indeed, if we replace the microscopic problem inside
each box with the homogenized problem with Neumann boundary conditions, the $O(\frac{\epsilon}{\Delta t})$-term
vanishes.  The result is shown in table \ref{tab:diff-fd-gt-2-reac}.
\begin{table}
\begin{center}
\begin{tabular}{|l||r|r||r|r|} \hline
   & \multicolumn{2}{|c||}{$h=0.005$}
   & \multicolumn{2}{|c|}{$h=0.01$}\\
   \hline
   &  error \phantom{blah} & ratio & error \phantom{blah} & ratio \\ \hline
  $\Delta t=4\cdot 10^{-3}$ & $1.0812\cdot 10^{-4}$ &     &   $1.0982\cdot  10^{-4}$ &
  \\\hline
  $\Delta t=2\cdot 10^{-3}$   & $4.9371\cdot 10^{-5}$ &  2.19    &  $5.1135\cdot  10^{-5}$ & 2.15
  \\\hline
  $\Delta t=1\cdot 10^{-3}$  & $2.1459\cdot 10^{-5}$ & 2.3     & $2.3411\cdot   10^{-5}$ &  2.18
  \\\hline
  $\Delta t=5\cdot 10^{-4}$   & $8.1325\cdot 10^{-5}$ & 2.63   &   $1.0492\cdot  10^{-5}$ & 2.23 \\
  \hline
\end{tabular}
\caption{Difference between the gap-tooth scheme for
(\ref{eq:ex1_fine}) and a finite difference scheme for
(\ref{eq:ex1_coarse}) for order $k=2$ and $\Delta x=0.05$ at time
$t=2\cdot 10^{-2}$.\label{tab:diff-fd-gt-2-reac}}
\end{center}
\end{table}
We see that the decrease of convergence speed indeed disappears.
The observed ratios are slightly larger than 2,
due to the interference with the $O(h^2)$ term due to averaging,
which is opposite in sign.

\subsection{Rough non-periodic (random) diffusion \label{sec:5.3}}

With this example, we illustrate that the scheme can also be used
to simulate systems with rough coefficients,
which are correlated on a length scale $\epsilon$.
We use the same example as Abdulle and E \cite{AbdE03},
who constructed it to demonstrate
the behaviour of the heterogeneous multi-scale method.
We first take a uniformly distributed random signal $s(x)$ in
$[0.1,1.1]$, with $x\in [0,1]$.  We discretize the interval $[0,1]$ in $N$
equidistant points $x_i$, and define
a correlation kernel $g^{\epsilon}(x)$, such that
\begin{displaymath}
g^{\epsilon}(0)=\frac{1}{\epsilon}, \qquad g^{\epsilon}(x)=0
\text{ if } x \notin (-\frac{\epsilon}{2},\frac{\epsilon}{2}),
\qquad \int_{-\frac{\epsilon}{2}}^{\frac{\epsilon}{2}}g^{\epsilon}(x)\d x=1.
\end{displaymath}
Here, we choose
$g^{\epsilon}(x)=\frac{1}{\epsilon}(1-\sin(\frac{2\pi x}{\epsilon}))$,
$x \in (-\frac{\epsilon}{2},\frac{\epsilon}{2})$.
We then define the rough diffusion coefficient in the discretization points as
\begin{displaymath}
a^{\epsilon}(x_i)=(g^{\epsilon}*s)(x_i)=
\int_{-\frac{\epsilon}{2}}^{\frac{\epsilon}{2}}g^{\epsilon}(x_i-\xi)s(\xi)\d\xi.
\end{displaymath}
We then consider the diffusion equation
\begin{displaymath}
\frac{\partial}{\partial t}u_{\epsilon}(x,t)=
\frac{\partial}{\partial x}\left(a^{\epsilon}(x)\frac{\partial}{\partial x}u_{\epsilon}(x,t)\right),
\end{displaymath}
with $a^{\epsilon}(x)$ constructed as above.
Note that this is a deterministic problem once the
diffusion coefficient is obtained.  We consider only one realization of the diffusion coefficient field.
Here, we can approximate the effective behaviour by averaging in space; in \cite{RunTheoKevr02}, one
had to average over many (shifted) initial conditions.

\begin{figure}
\begin{center}
\subfigure{\includegraphics[scale=0.35]{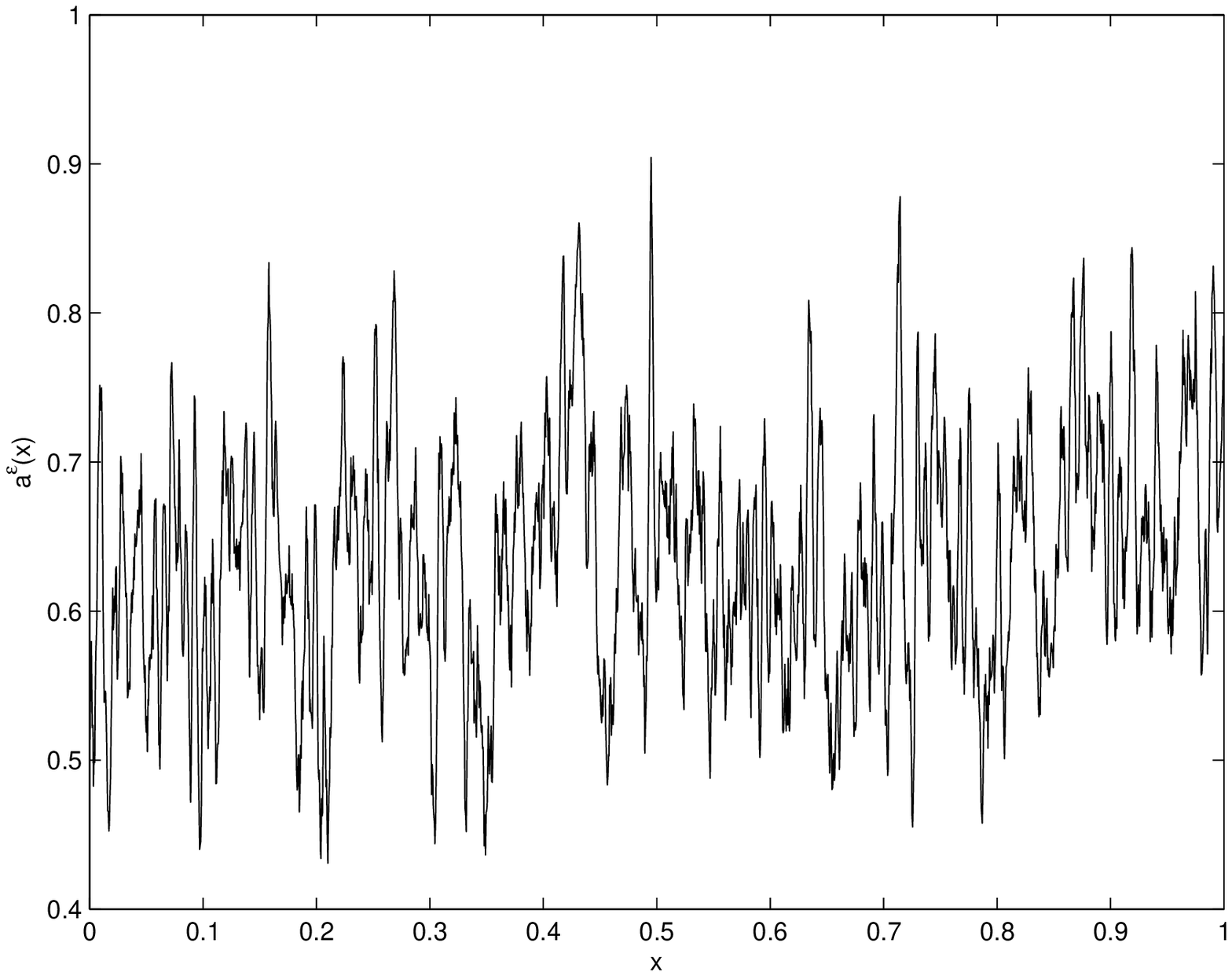}}
\subfigure{\includegraphics[scale=0.35]{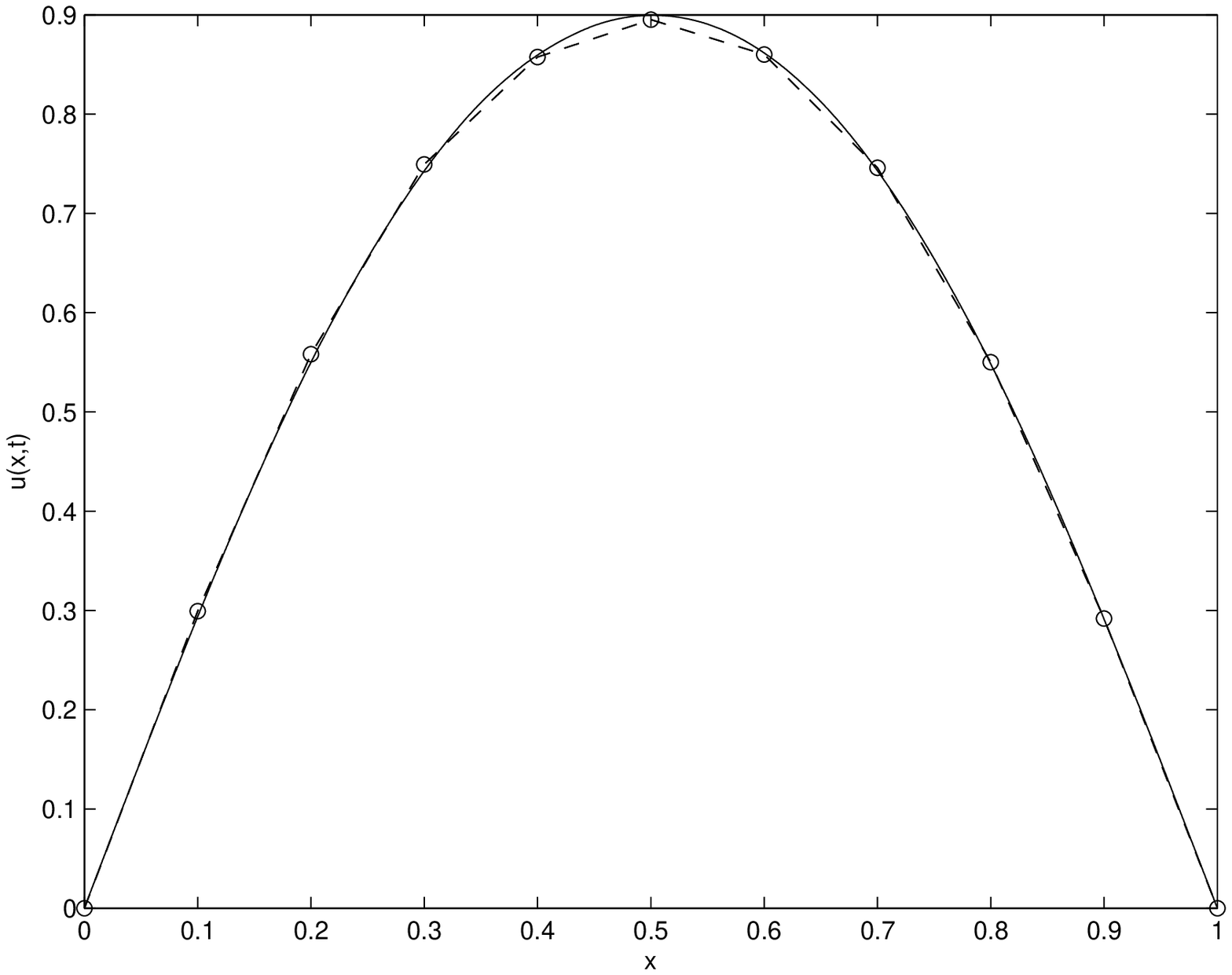}}
\caption{\label{fig:random_diff}Left: Random correlated diffusion coefficient $a^{\epsilon}(x)$.  Right:
Reference solution (with $N=20001$, $\Delta t=1\cdot 10^{-5}$ with an implicit Euler time-step) and gap-tooth solution
(with $\Delta x=0.1$, $\Delta t=2\cdot 10^{-3}$), at time $t=2\cdot 10^{-2}$.}
\end{center}
\end{figure}

We compute the solution with $N=20001$ and $\epsilon=1\cdot 10^{-3}$,
and we compare this to a gap-tooth solution with $h=5\cdot 10^{-3}$.  As an initial condition,
we take $u_{\epsilon}(x,0)=1-4(x-0.5)^2$, with homogeneous Dirichlet boundary conditions.
Figure \ref{fig:random_diff} shows the diffusion coefficient and the reference and gap-tooth solutions.
We compute the solution for this problem with an increasing number of boxes,
where $\nu=0.2$ is fixed and $\Delta t=\nu\Delta x^2$.
The error is shown in table \ref{tab:diff-fd-gt-1-rand}.
\begin{table}\begin{center}
\begin{tabular}{|c|r|}
\hline
& error\phantom{blah}\\
\hline
$\Delta x=2\cdot 10^{-1}$ & $1.0283\cdot 10^{-2}$\\
$\Delta x=1\cdot 10^{-1}$ & $8.1064\cdot 10^{-3}$\\
$\Delta x=5\cdot 10^{-2}$ & $7.4622\cdot 10^{-3}$\\
$\Delta x=2.5\cdot 10^{-2}$ & $1.4028\cdot 10^{-3}$\\
\hline
\end{tabular}
\caption{\label{tab:diff-fd-gt-1-rand}Difference between the gap-tooth scheme with $h=5\cdot 10^{-3}$
and the reference solution at time $t=2\cdot 10^{-2}$ for increasing level of discretization.}
\end{center}
\end{table}
We see that the error decreases when more boxes are inserted.  Note that, due to the
roughness of the signal, it is difficult to draw conclusions on the convergence rate, and to
determine good parameter values for the gap-tooth scheme.
E.g.~the length over which the gradient is averaged
at the end points of each box is no longer uniquely defined,
since the small scale $\epsilon$ is only a correlation length.

\section{Avoiding the algebraic constraint \label{sec:dispersion}}

We recall that the gap-tooth scheme, as presented above, performs the simulations
inside each box using an algebraic constraint,
ensuring that the initial macroscopic gradient
is preserved at the boundary of each box over the time-step $\Delta t$.
If our goal is to accelerate time-integration using an existing microscopic code,
this constraint may require us to alter this code, so as to impose this
macroscopically-inspired constraint.
This may be impractical (e.g.\ if the macroscopic gradient has to be estimated),
undesirable
(e.g.\ if the development of the code is expensive and time-consuming) or even
impossible (e.g.\ if the microscopic code is a \emph{legacy code}).

Generally, a given microscopic code
allows us to run with a set of pre-defined boundary
conditions.
It is highly non-trivial to impose macroscopically inspired boundary
conditions on such microscopic codes, see e.g.\ \cite{LiYip98} for a control-based
strategy.
This can be circumvented by introducing buffer regions at the boundary
of each small box, which shield the dynamics within the computational
domain of interest from boundary effects during a short time interval.  One then uses
 the microscopic code with its \emph{built-in} boundary conditions \cite{SamKevrRoo03}.

\subsection{The gap-tooth scheme with buffers}

\begin{figure}
\begin{center}
\includegraphics[scale=0.5]{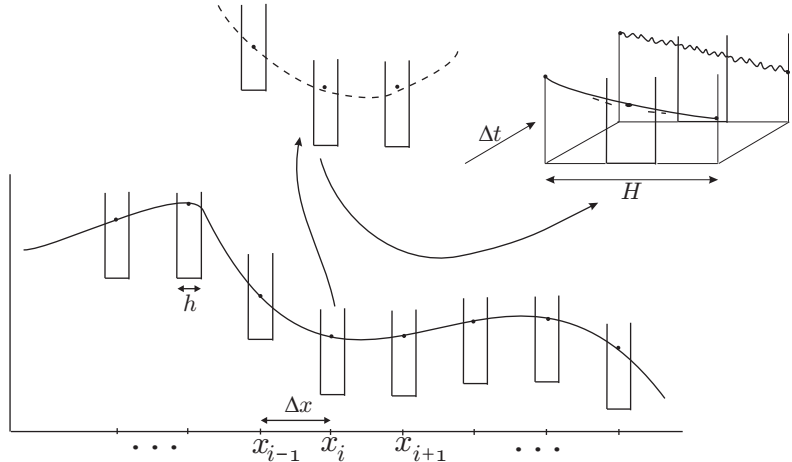}
\caption{\label{fig:schematic_buffer}A schematic representation of
the gap-tooth scheme with buffer boxes.  The simulation is done in the box
of size $H$, whereas for restriction, only information inside the inner box of size
$h$ is used.}
\end{center}
\end{figure}

We note that, for a correct simulation,
the only crucial issue is that the detailed system in each box should evolve
\emph{as if it were embedded in a larger domain}.
This can be effectively accomplished by introducing a larger box of size $H>>h$
around each
macroscopic mesh point, but still only use (for macro-purposes) the evolution over
the smaller, ``inner" box.
This is illustrated
in figure \ref{fig:schematic_buffer}.
Lifting and evolution (using \emph{arbitrary} outer boundary
conditions) are performed in the larger box; yet the restriction is done by
taking the average of the solution over the inner, small box.
The goal of the additional computational domains, the \emph{buffers}, is to
buffer the solution inside the small box from outer boundary effects.
This can be accomplished over \emph{short enough} times, provided the buffers are \emph{large enough};
analyzing the method is tantamount to making these statements quantitative.

The idea of using a buffer region was also used in the multi-scale finite element method
(oversampling) of Hou \cite{HouWu97}
to eliminate the boundary layer effect; also Hadjiconstantinou makes use of overlap
regions to couple a particle simulator with a continuum code \cite{Hadji99}.
If the microscopic code allows a choice of
different types of ``outer" microscopic boundary conditions,
selecting the size of the buffer may
also depend on this choice.

\subsection{Damping factors}

Here, we show that we can study the gap-tooth scheme (with buffers)
through its numerically obtained damping factors,
i.e.~by estimating the eigenvalues of its linearization.
Integration with nearby coarse initial conditions is used to
estimate matrix-vector products of the linearization of the coarse time-$\Delta t$ map
with known perturbation vectors; these are integrated in matrix-free iterative
methods such as Arnoldi eigensolvers.
For the diffusion homogenization problem (\ref{eq:ex1_fine}),
we show that the eigenvalues of the gap-tooth scheme are approximately
the same as those of the corresponding finite difference scheme for
(\ref{eq:ex1_coarse}).  When we impose
Dirichlet boundary conditions at the boundary of the buffers, we show
that the scheme converges to the standard gap-tooth scheme for increasing
buffer size.

Convergence results are typically established by proving
 consistency and stability.
If one can prove that the error in each time step can be
 made arbitrarily small by
refining the spatial and temporal mesh size, and that an
 error made at time $t_n$ does
not get amplified in future time-steps, one has proved
convergence.  This requires the solution operator to be
stable as well. In the absence of explicit formulas, one can examine
 the damping factors of the
time-stepper.  If, for decreasing mesh sizes, all
(finitely many) eigenvalues
and eigenfunctions of the
time-stepper converge to the
dominant eigenvalues and eigenfunctions of the
time evolution operator, one expects the solution
of the scheme to converge to the true solution of the evolution problem.

Consider equation (\ref{eq:ex1_coarse}) with Dirichlet boundary
conditions $u(0,t)=0$ and $u(1,t)$, and denote its solution at
time $t$ by the time evolution operator
\begin{equation}\label{eq:evolution_operator}
u(x,t)=s(u_0(x);t),
\end{equation}
We know that
\begin{displaymath}
s(\sin(m\pi x);t)=\e^{-(m\pi)^2 t} \sin(m\pi x), \qquad m\in
\mathbb{N}.
\end{displaymath}
Therefore, if we consider the time evolution operator over a fixed
time $\bar{t}$, $s(\cdot,\bar{t})$, then this operator has
eigenfunctions $\sin(m\pi x)$, with resp.\ eigenvalues
\begin{equation}\label{eq:dispersion_relation}
\lambda_m=\e^{-{(m\pi)}^2\bar{t}}.
\end{equation}
A good (finite
difference) scheme approximates well all
eigenvalues whose eigenfunctions can be
represented on the given mesh. We choose
$\bar{t}$ as a multiple of $\Delta t$ for convenience.

Since the operator defined in (\ref{eq:evolution_operator})
 is linear, the
numerical time integration is equivalent to a
matrix-vector product.  Therefore,
we can compute the eigenvalues using matrix-free
linear algebra techniques, even for the gap-tooth scheme,
for which it might not even be
possible to obtain a closed
expression for the matrix.
Note that this idea is general; here we use it as a tool to study the effect
of the buffer size on convergence of the gap-tooth scheme.
However, although this analysis gives us an indication about the
quality of the scheme, it is by no means a proof of
convergence.

\subsection{Numerical results}

\begin{figure}
\centering
\subfigure{\includegraphics[scale=0.3]{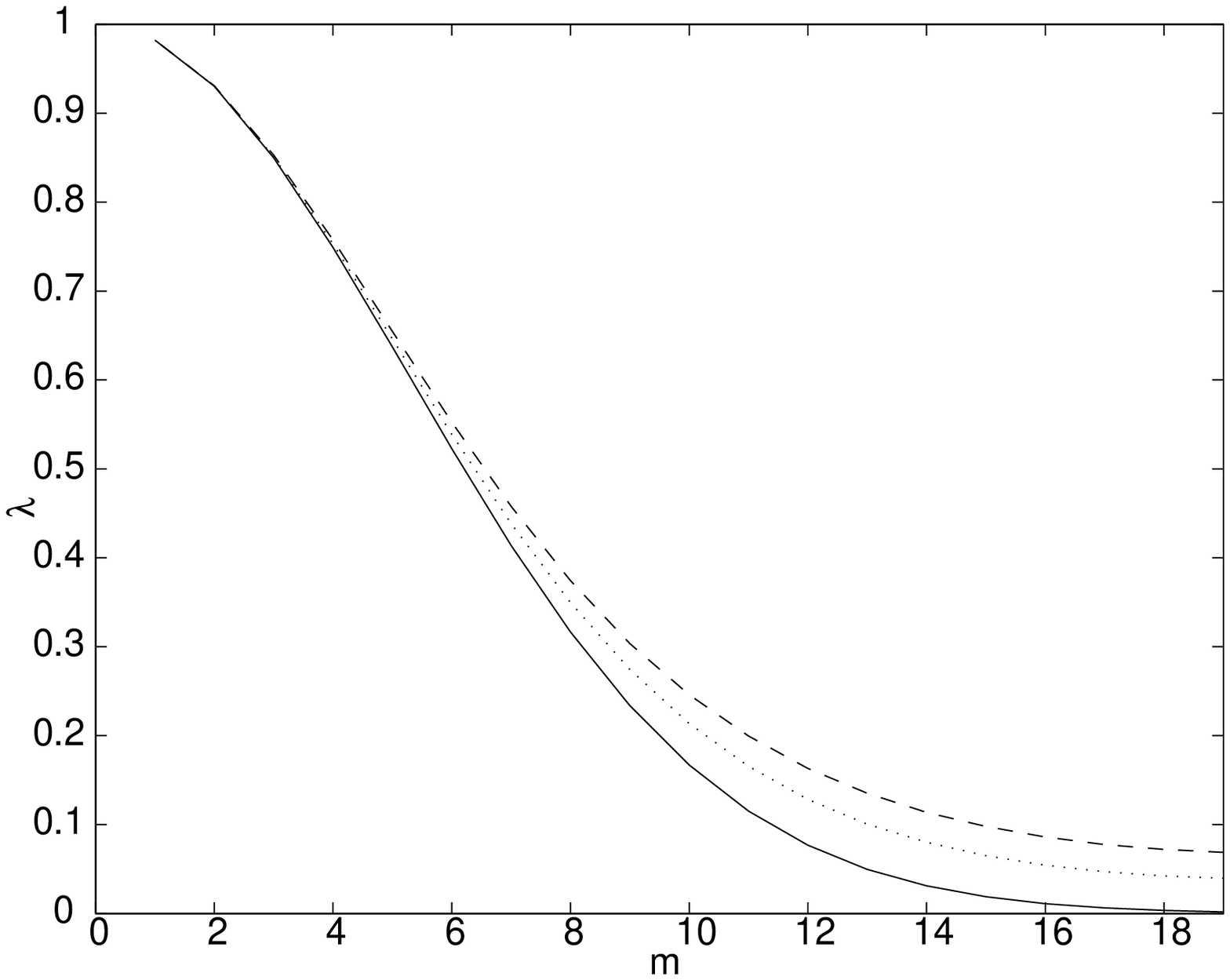}}
\subfigure{\includegraphics[scale=0.3]{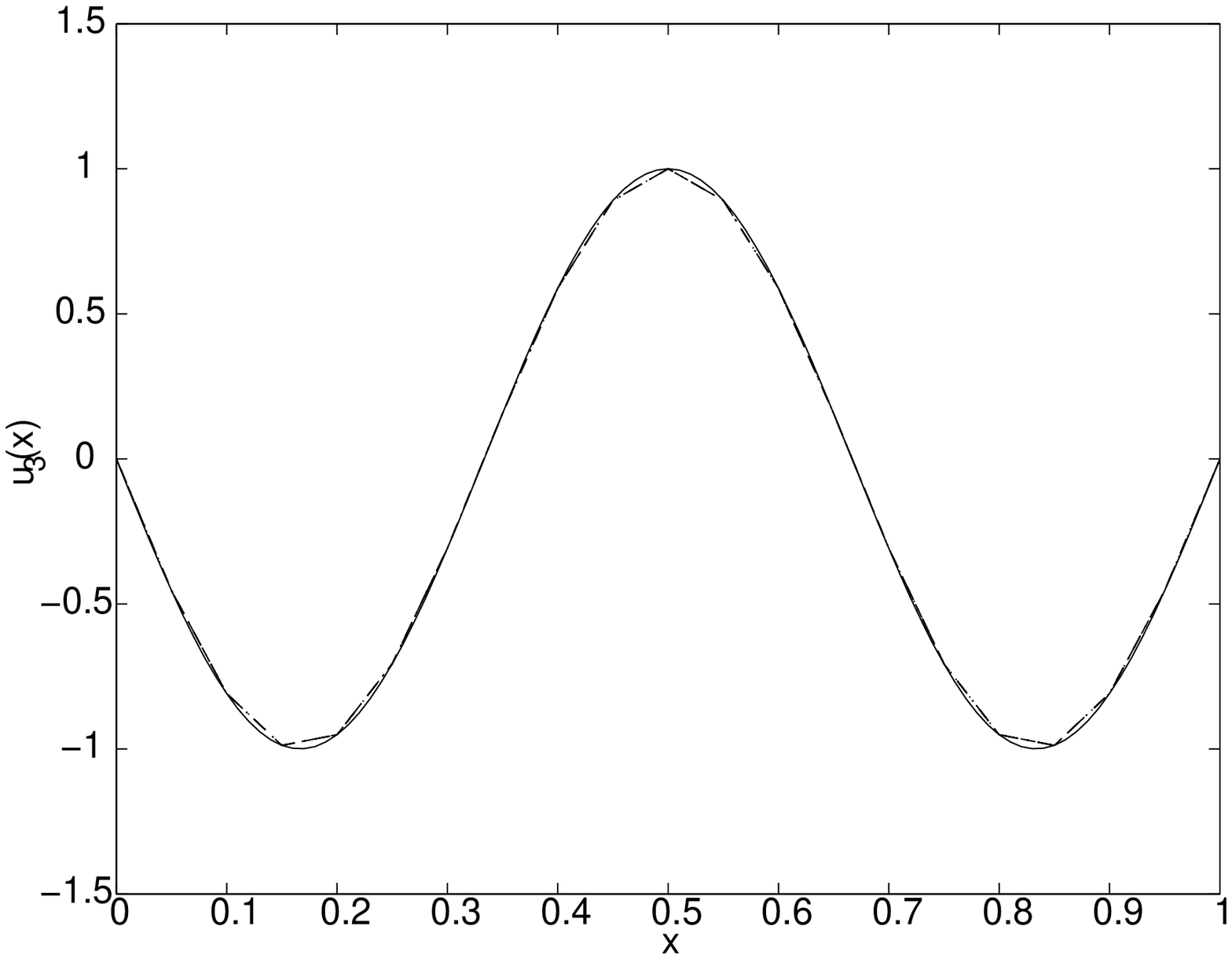}}
\caption{\label{fig:eigval_neu}Comparison between
the damping factors (left) and the eigenfunction
corresponding to eigenvalue $\lambda_3$ (right) of the
exact solution (full line), the finite difference
approximation (dashed) and the gap-tooth scheme (dotted).
The eigenfunction of the gap-tooth scheme is indistinguishable
of the finite difference eigenfunction.}
\end{figure}

\begin{figure}
\centering
\subfigure{\includegraphics[scale=0.3]{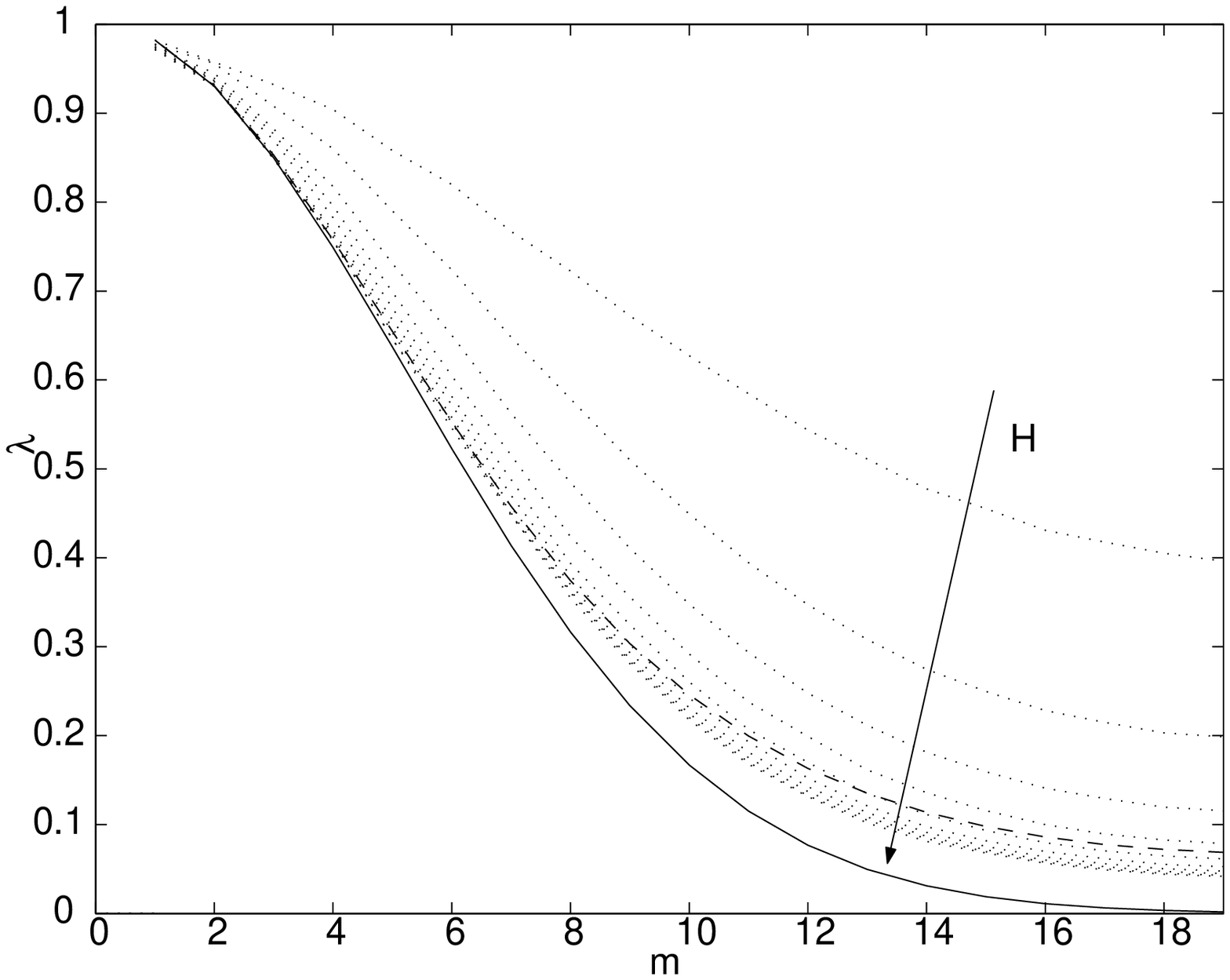}}
\subfigure{\includegraphics[scale=0.3]{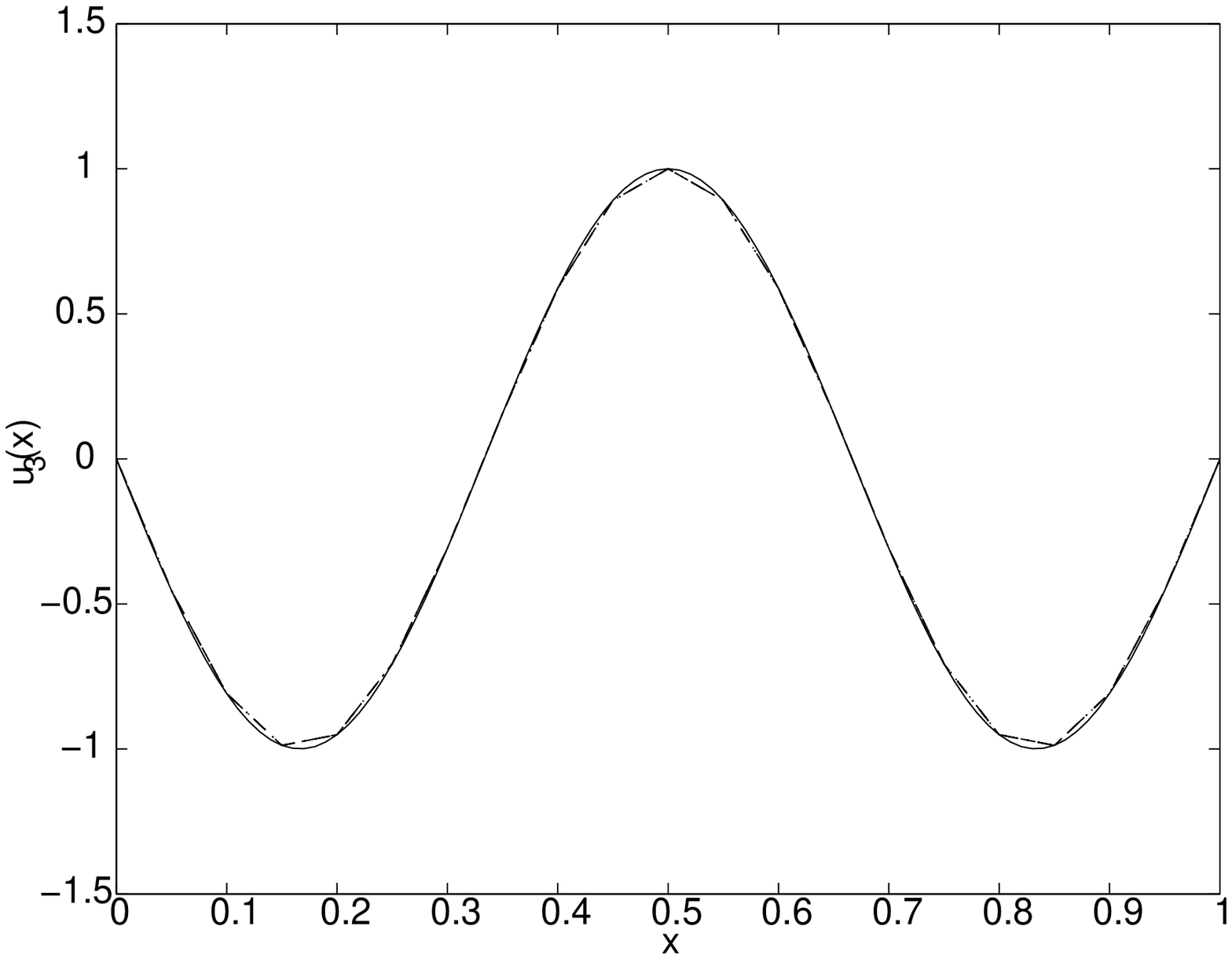}}
\caption{\label{fig:eigval_dir}Comparison between the damping factors (left) and
the eigenfunction corresponding to the eigenvalue $\lambda_3$ (right) of
the exact solution (full line), the finite difference scheme (dashed) and
the gap-tooth scheme with buffers (dash-dotted lines) for increasing buffer sizes
$H=2\cdot 10^{-2},3\cdot 10^{-2}\ldots,1\cdot 10^{-1}$.}
\end{figure}

We illustrate this idea by computing the eigenvalues of the
gap-tooth scheme of order $k=2$, applied to (\ref{eq:ex1_fine}).  In this case,
we know from section \ref{sec:convergence} that these eigenvalues
should approximate
the eigenvalues of a finite difference scheme on the same mesh.
 As method parameters,
we choose $\Delta x=0.05$,
$h=5\cdot 10^{-3}$,
$\Delta t=2.5\cdot 10^{-4}$ for a time horizon
$\bar{t}=4\cdot 10^{-3}$, which corresponds to 16 gap-tooth steps.
Inside each box,
we use a finite difference scheme of order
$2$ with $\delta x=1\cdot 10^{-5}$ and an implicit
Euler time-step of $5\cdot 10^{-7}$.
We compare
these eigenvalues to those of the finite difference
scheme with $\Delta x=0.05$ and
$\Delta t=2.5\cdot 10^{-4}$, and with the dominant
eigenvalues of the reference solution
(a finite difference approximation with
$\Delta x=1\cdot 10^{-5}$, $\Delta t=5\cdot 10^{-7}$ and implicit Euler time-stepping).
The result is shown in figure \ref{fig:eigval_neu}.
We now introduce a buffer region of size $H$, and we impose
Dirichlet boundary conditions at the outer boundary
of the buffer region.  Lifting is done in identically the same way as
for the gap-tooth scheme without buffers; we only use (\ref{eq:lifting})
as the initial condition in the larger box $[x_i-\frac{H}{2},x_i+\frac{H}{2}]$.
We compare the eigenvalues again with the
equivalent finite difference scheme and
the exact solution, for increasing sizes of the buffer box $H$.
Figure \ref{fig:eigval_dir} shows that, as $H$ increases, the eigenvalues
of the scheme
converge to those of the original gap-tooth scheme.   We see that,
in this case, we would need a buffer of size $H=4\cdot 10^{-2}$, i.e.\ 80\%
of the original domain, for a good approximation of the damping factors.
It is possible to decrease the buffer size by decreasing $\Delta t$, which results
in more re-initializations.

\section{Conclusions\label{sec:conclusions}}
We described the gap-tooth scheme for the numerical simulation of
multi-scale problems.  This scheme simulates the macroscopic behaviour
over a macroscopic domain when only a microscopic model is explicitly
available.  We analyzed the convergence of this
scheme for a parabolic homogenization problem with non-linear
reaction.

We
showed that our method approximates a finite difference scheme of arbitrary (even) order
for the homogenized equation when we
appropriately constrain the microscopic problem in the boxes, and
illustrated this theoretical result with numerical tests on several
model problems.
Our analysis revealed that the presence of microscopic scales, combined with
the requirement that the macroscopic gradient does not change over one gap-tooth
time-step $\Delta t$, introduces an error term that grows with decreasing $\Delta t$,
which is not optimal.

We also demonstrated that it is possible to obtain
a convergent scheme without constraining the microscopic code, by
introducing buffers that \emph{shield} over relatively short time intervals
the dynamics inside each box from boundary effects.
 It is possible, even without
analytic formulas, to study the properties
of the gap-tooth scheme and generalizations through the damping
factors of the
resulting coarse time-$\Delta t$ map.
In a forthcoming paper, we will use these damping factors
to study the the trade-off between
the effort required to impose a particular type of boundary conditions
 and the efficiency gain
due to smaller buffer sizes and/or longer possible time-steps before
reinitialization.

The time-stepper as constructed in this paper will allow us to perform simulations
of the effective behaviour of a microscopic system over macroscopic space and macroscopic time
(when combined with projective integration), or to perform tasks as bifurcation analysis or
coarse control (when coupled to time-stepper based bifurcation codes).

\section*{Acknowledgments}
The authors thank Sabine Attinger and Petros Koumoutsakos for organizing
the Summer School in Multi-scale Modeling and Simulation in Lugano, and the participants
for many fruitful discussions.
Giovanni Samaey is a Research Assistant of the Fund for Scientific Research - Flanders.
This work has been partially supported
by grant IUAP/V/22 and by the Fund of Scientific Research
through Research Project G.0130.03 (GS, DR),
and an NSF/ITR grant and AFOSR Dynamics and Control, Dr. B. King (IGK).

\bibliographystyle{plain}

\end{document}